\documentclass [11pt]{article}
\usepackage{amsfonts}
\usepackage{graphicx}
\usepackage{amsmath,amssymb, amsfonts, dsfont}
\usepackage{latexsym}
\usepackage[normalem]{ulem}
\usepackage{mathrsfs}
\usepackage{bbold}
\usepackage{color}
\usepackage{slashed}
\usepackage{cancel}
\usepackage{soul} 
\usepackage{setspace} 
\usepackage{comment}
\usepackage{booktabs}
\usepackage{multirow}
\usepackage{tikz-feynman}
\tikzfeynmanset{compat=1.0.0}
\usepackage{cite}
\usepackage[small]{caption}
\usepackage{epsfig}
\usepackage{slashed}
\usepackage{epstopdf}
\usepackage{latexsym}
\usepackage{graphicx}
\usepackage{subcaption}
\usepackage{tikz}
\usepackage{circuitikz}
\usetikzlibrary{decorations.pathmorphing}
\usetikzlibrary{decorations.markings}
\pgfdeclaredecoration{complete sines}{initial}
{
    \state{initial}[
        width=+0pt,
        next state=sine,
        persistent precomputation={\pgfmathsetmacro\matchinglength{
            \pgfdecoratedinputsegmentlength / int(\pgfdecoratedinputsegmentlength/\pgfdecorationsegmentlength)}
            \setlength{\pgfdecorationsegmentlength}{\matchinglength pt}
        }] {}
    \state{sine}[width=\pgfdecorationsegmentlength]{
        \pgfpathsine{\pgfpoint{0.25\pgfdecorationsegmentlength}{0.5\pgfdecorationsegmentamplitude}}
        \pgfpathcosine{\pgfpoint{0.25\pgfdecorationsegmentlength}{-0.5\pgfdecorationsegmentamplitude}}
        \pgfpathsine{\pgfpoint{0.25\pgfdecorationsegmentlength}{-0.5\pgfdecorationsegmentamplitude}}
        \pgfpathcosine{\pgfpoint{0.25\pgfdecorationsegmentlength}{0.5\pgfdecorationsegmentamplitude}}
}
    \state{final}{}
}

\usepackage{rotating}
\usepackage{bbm}
\usepackage{bbold}
\usepackage{enumitem}

\usepackage{empheq}

\definecolor{myblue}{rgb}{.8, .8, 1}
\newcommand*\mybluebox[1]{%
\colorbox{myblue}{\hspace{0.5em}#1\hspace{0.5em}}}

\definecolor{red}{rgb}{1,0,0}
\definecolor{extragreen}{RGB}{0,210,65}

\usetikzlibrary{positioning}
\tikzset{
    mybluenode/.style={
        draw=black, circle, minimum width=2cm, inner sep=0pt
        },
    myblacknode/.style={
        circle, inner sep=1pt, fill=black
        },
    }
\newcommand*\de{\partial}

\newcommand*\al{\alpha}

\newcommand*\ga{\gamma}
\newcommand*\di{\delta}

\newcommand*\la{\lambda}
\newcommand*\si{\sigma}

\newcommand*\ZZ{\mathbb{Z}}

\newcommand*\calD{\mathcal{D}}

\newcommand*\calO{\mathcal{O}}

\newcommand*\Tr{\mathop{\rm tr}}

\newcommand*\intads{\int}
\makeatletter
\def\section{\@startsection {section}{1}{\z@}{-3.5ex plus -1ex minus
 -.2ex}{2.3ex plus .2ex}{\large\bf}}
\def\subsection{\@startsection{subsection}{2}{\z@}{-3.25ex plus -1ex
minus -.2ex}{1.5ex plus .2ex}{\normalsize\bf}}
\makeatother
\makeatletter

\@addtoreset{equation}{section}

\makeatother

\newcommand{\bea}{\begin{equation} \begin{aligned}} \newcommand{\eea}{\end{aligned} \end{equation}}
\def\be{\begin{equation}} \def\ee{\end{equation}} 
\def\nn{\nonumber}

\usepackage[colorlinks,linkcolor=black,citecolor=blue,urlcolor=blue,linktocpage,pagebackref]{hyperref}

\allowdisplaybreaks

\begin{document}

\thispagestyle{empty}

\begin{center}

	\vspace*{-.6cm}

	\begin{center}

		\vspace*{1.1cm}

	{\centering \Large\textbf{QCD in AdS}}

	\end{center}

	\vspace{0.8cm}
	{\bf Riccardo Ciccone$^{a,b}$, Fabiana De Cesare$^{c,d}$, Lorenzo Di Pietro$^{e,f}$, Marco Serone$^{f,g}$}

	\vspace{1.cm}

 ${}^a\!\!$
	{\em Department of Physics and Haifa Research Center for Theoretical\\
Physics and Astrophysics, University of Haifa, 31905 Haifa, Israel}

	\vspace{.3cm}
 ${}^b\!\!$
	{\em Department of Physics, Technion,\\ Israel Institute of Technology, 32000 Haifa, Israel}

	\vspace{.3cm}
  ${}^c\!\!$
	{\em Institut des Hautes \'Etudes Scientifiques, 91440 Bures-sur-Yvette, France}
	
    \vspace{.3cm}
    ${}^d\!\!$
	{\em  INFN, Sezione di Torino, and Department of Physics, University of Turin,\\
Via P. Giuria 1, 10125, Turin, Italy}

     \vspace{.3cm}
    ${}^e\!\!$
	{\em  Dipartimento di Fisica, Universit\`a di Trieste, \\ Strada Costiera 11, I-34151 Trieste, Italy}

	\vspace{.3cm}

	${}^f\!\!$
	{\em INFN, Sezione di Trieste, Via Valerio 2, I-34127 Trieste, Italy}

	\vspace{.3cm}
${}^g\!\!$
	{\em SISSA, Via Bonomea 265, I-34136 Trieste, Italy}

\end{center}

\vspace{1cm}

\centerline{\bf Abstract}
\vspace{2 mm}
\begin{quote}
We study QCD on AdS space with scalars or fermions in the fundamental representation, extending earlier results on pure Yang-Mills theory. In the latter, the Dirichlet boundary condition is conjectured to disappear via merger and annihilation, as signaled by the lightest scalar singlet operator approaching marginality as the coupling increases. With matter, there are two candidate operators for this mechanism. We compute their one-loop anomalous dimensions via broken conformal Ward identities and Witten diagrams. In the confining phase, with Dirichlet (Neumann) boundary condition, their anomalous dimensions are negative (positive), consistent with the disappearance (persistence) of the associated boundary CFT in the flat-space limit.
In the conformal window, one of these operators becomes the displacement operator of the IR CFT, as signaled by the vanishing of its one-loop anomalous dimension in the perturbative Banks-Zaks regime. 
Possible scenarios for the lower edge of the conformal window are discussed. 
Finally, we consider general boundary conditions on fermions and discuss their relation to chiral symmetry breaking in flat space.
\end{quote}

\newpage

\tableofcontents

\newpage 
\section{Introduction}

In this paper we study non-abelian gauge theories in four-dimensional Anti-de Sitter (AdS) space, extending the methods applied in \cite{Ciccone:2024guw} to pure $SU(n_c)$ Yang-Mills (YM) theory to the case with fundamental matter fields, i.e. quantum chromo-dynamics (QCD), either fermionic or bosonic. 

The idea of considering non-abelian gauge theories in non-trivial spaces, to shed light on confinement and other properties of the flat-space theory,
is of course not new and dates back at least to  \cite{tHooft:1979rtg,Eguchi:1982nm}.
Most analyses focused on toroidal compactifications, with the idea that at small volume one can study the spectrum perturbatively and then take the large volume limit to extrapolate to the flat-space spectrum; see \cite{Luscher:1982ma} for a concrete early study along these lines. 
In order to avoid a confinement-deconfinement phase transition occurring as the volume is varied, non-trivial holonomies around toroidal spaces have been proposed, see e.g.\cite{Kovtun:2007py,Tanizaki:2022ngt}. Anti-de Sitter space offers us a new perspective and advantages with respect to, e.g., toroidal compactifications. AdS is a maximally symmetric space with a conformal boundary at infinity.
The continuity of the flat-space limit in AdS depends on the choice of the boundary condition (bc), in close analogy with the role of holonomies in toroidal geometries. The role of volume in compact spaces is played by the AdS radius $L$, which combines with the dynamically generated QCD scale $\Lambda$ to form a dimensionless parameter $\Lambda L$. We can then set up a perturbative expansion for $\Lambda L\ll 1$, and interpolate between the UV and IR regimes of the theory when the appropriate bc is chosen. 
Notably, the presence of a conformal symmetry at the boundary represents an important step forward, as we understand and control conformal theories much better than ordinary QFTs. The non-local nature of the boundary theory does not represent a real obstacle, as conformal theories can be approached axiomatically, like in the conformal bootstrap \cite{Poland:2018epd}. In fact, approaching flat space from AdS is a promising direction more in general, 
 e.g. as an alternative way to axiomatically define flat-space S-matrix amplitudes \cite{Paulos:2016fap} and 
 to better understand their analytic properties, see e.g. \cite{vanRees:2022zmr}. 

Non-abelian gauge theories on AdS, first considered in \cite{Callan:1989em}, 
have been later revisited in \cite{Aharony:2012jf}, where it was emphasized that the condition of color confinement in flat space requires that the operators of the boundary conformal theory do not carry charges under the gauge group $G$.
On the other hand, if one imposes the Dirichlet bc for the gauge fields one gets precisely such a $G$ symmetry at the boundary, and also local boundary operators charged under it. This apparent contradiction leads to the idea that the Dirichlet bc only exists in the perturbative regime of small $\Lambda L$, and it disappears at some critical value. The disappearance of the Dirichlet bc is therefore a clear manifestation of confinement from the point of view of the boundary.

But how does a boundary condition disappear? A natural possibility is that this happens through the general CFT phenomenon of merger and annihilation \cite{Kaplan:2009kr, Gorbenko:2018ncu, Gorbenko:2018dtm}, which was considered in the context of QFT in AdS in \cite{Hogervorst:2021spa, Lauria:2023uca}. This explanation for the disappearance of the Dirichlet bc of four-dimensional YM theory was proposed in \cite{Copetti:2023sya} and perturbative evidence in its favor was presented in \cite{Ciccone:2024guw}.\footnote{Alternative proposals discussed in \cite{Aharony:2012jf} for the disappearance of the Dirichlet bc include the possibility that the primary state associated to the global current operator becomes null (decoupling), 
or that a bulk Higgs mechanism of some kind occurs, signaled by a boundary charged operator becoming marginal (Higgsing). 
Upcoming conformal bootstrap studies rule out the former, and disfavor the latter with respect to the merger and annihilation proposal \cite{DiPietro:2025ym}.}
In particular, it was found that the lowest scalar singlet operator, which in the free limit is the displacement operator of the free boundary CFT (BCFT), of 
dimension $\Delta =4$, gets a negative one-loop anomalous dimension. This suggests that at some finite $\Lambda L$ it reaches the marginal value $\Delta =3$ and triggers merger and annihilation with a second boundary condition, dubbed Dirichlet$^*$ in \cite{Ciccone:2024guw}. Making the alternative choice of Neumann bc, the anomalous dimension of the lowest scalar singlet, also of dimension $\Delta =4$ in the free limit, was instead found to be positive, consistently with the fact that all the boundary operators are color-neutral in this case, and there is no reason to expect a discontinuity in the flat-space limit.
Besides \cite{Callan:1989em, Aharony:2012jf, Ciccone:2024guw}, some previous studies of the dynamics of gauge theories in Anti-de Sitter space are \cite{Allen1986, DHoker:1998bqu, DHoker:1999bve, DHoker:1999mqo, Rattazzi:2009ux,  Aharony:2010ay,  Paulos:2011ie,  Raju:2012zs, Aharony:2015zea, Albayrak:2020bso, Albayrak:2020fyp,   Caron-Huot:2021kjy, Alday:2021odx,   Behan:2023fqq, Albayrak:2023kfk,  Ankur:2023lum,  Albayrak:2024ddg,  Ghodsi:2024jxe,  Bason:2025zpy,  Gabai:2025hwf, Sleight:2025dmt}.

In the theories with fundamental matter, which are the topic of the present work, due to screening the notion of strict confinement does not apply. Here by ``strict confinement'' we mean the existence of a linearly rising potential between infinitely massive probe charges.\footnote{
In flat space, this condition is usually rephrased in terms of the Euclidean expectation value of large Wilson loops having area or perimeter law,  more formally reinterpreted as the statement that a one-form symmetry is spontaneously broken or unbroken \cite{Gaiotto:2014kfa}. Although in AdS one cannot use area vs perimeter as a diagnostic \cite{Callan:1989em}, it still makes sense to ask whether a one-form symmetry is spontaneously broken or not. However, both in AdS and in flat space, the presence of fundamental matter completely breaks the one-form symmetry of YM theory, making it impossible to define strict confinement in terms of the (lack of) spontaneous breaking of this symmetry.}
On the other hand, the weaker notion of ``color confinement'', 
meaning that all asymptotic states should be neutral under the gauge group $G$, still applies in the presence of matter. Only this weaker notion 
is needed to argue that the Dirichlet bc must disappear at large radius, so this is still a valid conclusion in QCD. Therefore, we are faced with a similar apparent contradiction as for pure YM theory, and we conclude that the Dirichlet bc for the gauge fields must disappear, while the Neumann bc can be continuous (at least assuming simple boundary conditions for the matter fields, that do not themselves introduce further discontinuities in the flat-space limit). It is then interesting to look at the scaling dimensions of the lowest-lying singlet operators in search of perturbative hints of merger and annihilation. 

An important novelty due to the inclusion of matter is that there are two low-lying singlet operators at the boundary, instead of the single one of YM theory. This is because at weak coupling the gluons and the matter fields give rise to two decoupled BCFTs, each with a displacement operator with scaling dimension $\Delta = 4$. We compute the anomalous dimensions of these operators at leading order, solving the associated mixing problem, both in the case of Dirichlet and Neumann bc. We do so by generalizing the analysis based on broken conformal Ward identities presented in 
\cite{Ciccone:2024guw} for a single bulk CFT deformed by a relevant operator, to the case of multiple decoupled CFTs. Using this method, only tree-level calculations are needed to compute the anomalous dimensions. In the case of the Dirichlet bc, we also perform an explicit check with a direct one-loop calculation in the bulk, finding agreement with the previous method.

The results that we obtain for the anomalous dimensions are in perfect agreement with the expectation of merger and annihilation for Dirichlet and continuity for Neumann: namely, both low-lying singlets get a negative anomalous dimension with Dirichlet, and a positive one with Neumann. The fact that the pattern found in pure YM persists in QCD, both fermionic and scalar, gives us more confidence in our interpretation of the perturbative results.  

Adding matter also gives us access to completely new physics, namely we can explore the cases in which QCD, rather than confining, flows to an interacting conformal field theory in the IR. This happens for instance in the theory with $n_f$ fundamental fermionic flavors, for sufficiently large number of flavors but still in the asymptotically free regime, i.e. $n_f^* \leq n_f < \tfrac{11}{2}n_c$.
The range of values for the number of flavors such that the theory flows to an interacting CFT is referred to as the  ``conformal window''. The value $n_f^*$, the onset of the conformal window, is a much studied but still unknown function of $n_c$. 

When QCD in the conformal window is studied in AdS, in the limit of large radius we are effectively studying a CFT in an AdS background, and we can use a Weyl rescaling to relate this setup to a BCFT in flat space with a boundary, i.e. a bulk CFT equipped with a conformally invariant boundary condition.  Therefore, the boundary spectrum in this case must include a displacement operator of dimension $\Delta = 4$ also at large radius. Checking the existence of this operator is a way to test quantitatively the flow to an IR CFT in AdS. In general, the IR CFT is reached at strong coupling; however, when one is parametrically close to the upper edge of the conformal window (which can be achieved either by taking $n_f$ as a continuous parameter, or more systematically in the Veneziano limit), the setup becomes calculable thanks to the Banks-Zaks perturbative expansion \cite{Caswell:1974gg, Banks:1981nn}. We perform the test in this perturbative limit, finding that indeed a displacement operator appears as expected in the IR spectrum. Notably, this operator is precisely a linear combination of the two low-lying operators mentioned above in the context of the confining phase. 
As a valuable consistency check supporting our result, we find this displacement operator with both Dirichlet and Neumann bc. We also obtain the scaling dimension in these BCFTs of the second, unprotected scalar singlet. We then speculate  on a possible relation between the disappearance of the Dirichlet BCFT in the conformal window and the end of the bulk conformal window, motivated by a hypothetical continuity of the boundary data across the transition. We note that, using this logic, the extrapolation of the one-loop data gives estimates for the end of the conformal window that are in good agreement with the existing methods.

An additional important physical phenomenon that happens in non-abelian gauge theories with fermionic matter, typically but not necessarily in conjunction with confinement, is the spontaneous breaking of a global flavor symmetry. The prototypical example is chiral symmetry breaking in four-dimensional QCD. In this paper we only touch upon the manifestation of chiral symmetry breaking from the point of view of the boundary CFT. We start from the observation of \cite{Rattazzi:2009ux} that chiral symmetry breaking happens even at weak coupling for gauge theories in AdS, and we discuss the possible patterns of symmetry breaking at weak coupling, showing that they depend on the parameters in the fermionic bc. We stress the interpretation in terms of spontaneous symmetry breaking in AdS, and we identify the AdS avatars of the pions in the protected tilt operators of the boundary CFT.\footnote{The existence of these protected operators in connection with chiral symmetry breaking was also discussed recently in \cite{Gabai:2025hwf}.} Finally, we discuss how, with Neumann bc, the weak coupling calculation of the fermion bilinear condensate can extrapolate to the flat-space non-perturbative result.

\paragraph{Outline} The rest of the paper is organized as follows. In section \ref{sec:displop} we generalize the method based on broken conformal Ward identities of \cite{Ciccone:2024guw} to the case of two or more decoupled CFT sectors in the UV, showing how to obtain the anomalous dimensions of the singlet scalar operators that arise from the displacement operators of the UV CFTs upon a relevant deformation. In section \ref{sec:fermions} and \ref{sec:scalars} we discuss the generalities of YM theories in AdS coupled to fermionic and scalar matter, respectively. For each case we solve the one-loop mixing problem for the displacement operators coming from the gauge and the matter sectors by using the method of section \ref{sec:displop} and, when the gauge fields satisfy the Dirichlet bc, also by a direct diagrammatic computation. In section \ref{sec:physical} we present the physical implications of our results: in the confining phase, we discuss the evidence in favor of merger and annihilation for the Dirichlet bc and in favor of continuity for the Neumann bc; we then discuss how our computations can detect the presence of bulk conformal phases, and discuss a speculative approach to estimate the lower edge of the conformal window. In section \ref{sec:chisb} we discuss a more general set of boundary conditions for the fermionic matter fields and its relation to chiral symmetry breaking. We conclude in section \ref{sec:outlook} with an outlook where we discuss possible future directions.

\section{Scaling dimensions from broken conformal Ward identities} 
\label{sec:displop}

A CFT in Euclidean AdS$_{d+1}$ can be mapped, via a Weyl rescaling to flat half-space, to a BCFT, which 
has a boundary scalar operator with scaling dimension exactly equal to $\Delta_{{\cal D}} = d+1$, the so-called displacement
operator ${\cal D}(\vec{x})$ \cite{McAvity:1993ue,McAvity:1995zd}. If a bulk deformation $O(x)$ is added, breaking (bulk) conformality, 
${\cal D}$ acquires an anomalous dimension, which is a function of the bulk coupling $\lambda$. 
As was shown in \cite{Ciccone:2024guw}, the violation of the Ward identity for the stress-energy tensor $T$ due to the breaking of (bulk) scaling gives a relation, to all orders in $\lambda$, between $\Delta_{\cal D}(\lambda)$ and
the bulk-to-boundary OPE coefficients $C_{T{\cal D}}(\lambda)$ and $C_{O{\cal D}}(\lambda)$. In the notable case in which the operator $O$ is classically marginal,
the case we will focus on in what follows, the relation also involves $\beta_\lambda(\lambda)$, the $\beta$-function of the coupling $\lambda$.

In this section, we generalize the results of section 4.1 of \cite{Ciccone:2024guw} to the case in which the bulk theory is a direct sum of $n$ 
free CFTs deformed by $m$ local operators $O_p$ $(p=1,\ldots,m)$, parametrized by coupling constants $\lambda_p$, which we denote collectively with the vector ${\boldsymbol{\lambda}}$. 
We consider AdS with metric
\begin{equation}\label{eq:poincmet}
ds^2 = L^2 \frac{d z^2 + d\vec{x}^2}{z^2}\,,
\end{equation}
where $\vec{x}$ denotes the directions parallel to the $d$-dimensional boundary, located at $z=0$.
In what follows and throughout the whole paper, depending on the context, we will write formulas either using Poincar\'e coordinates \eqref{eq:poincmet} or embedding space coordinates \cite{Costa:2011mg,Costa:2014kfa}. We will also set $L=1$ unless otherwise specified.

At ${\boldsymbol{\lambda}}=0$, we have $n$ decoupled stress-energy tensors $T^i$ and $n$ associated displacement operators ${\cal D}_i$ ($i=1,\ldots,n$) with scaling dimension $d+1$, defined by the boundary OPE of the stress tensor components perpendicular to the boundary:
\begin{equation}
    T^i_{zz} \underset{z\to 0}{\sim} b_{T^i \mathcal{D}^i}\, z^{d-1}\calD_i+\dots~.
\end{equation}
Their two-point functions read 
\begin{equation}\label{eq:Dis1}
\text{CFT:}\quad\langle T_{AB}^i(X) \mathcal{D}_j(P) \rangle =\delta_{ij}  C_{T^i\mathcal{D}^i} \frac{H_{AB}(X,P)}{(-2 P\cdot X)^{d+1}}\,,
\end{equation}
where here and in the following $A,B,\ldots$ are embedding space indices and 
\be
H_{AB} = \frac{G_{AC}(X)G_{BD}(X) P^C P^D}{(-2 P\cdot X)^2}- \frac{G_{AB}(X)}{4(d+1)}\,, \qquad  G_{AB}(X) = \eta_{AB}+X_A X_B\,,
\ee
is a tensor structure which is fixed by demanding conservation and traceless conditions on $T^i$.  We normalize the displacements ${\cal D}_i$ such that 
\begin{equation}\label{eq:Dis1a}
\text{CFT:}\quad\langle \mathcal{D}_i(\vec{x}_1) \mathcal{D}_j(\vec{x}_2) \rangle =  \frac{\delta_{ij}}{(\vec{x}_{12}^{\,2})^{d+1}}~.
\end{equation}
In this normalization, we have $b_{T^i \mathcal{D}^i} = C_{T^i\mathcal{D}^i}$.

When we add the bulk deformations
\begin{equation}\label{eq:SDeform}
S = \sum_{i=1}^n S_{\text{CFT}_i} + S_{\text{int}}\,,\qquad S_{\text{int}}= \sum_{p=1}^m \lambda_p \int dx\,O_p(x)\,,
\end{equation}
the individual $T^i_{\mu\nu}$ ($\mu,\nu$ being AdS indices) are no longer conserved. However, the total stress-energy tensor of the interacting theory, that we denote simply as $T_{\mu\nu}$,
remains conserved. When ${\boldsymbol{\lambda}}\neq 0$ typically bulk conformal symmetry is broken, resulting in the following nonzero trace of the stress tensor\footnote{We are ignoring possible curvature terms proportional to the identity operator because they end up giving a vanishing contribution to the anomalous dimension.}
\begin{equation}
    G^{AB}T_{AB}(X)= \sum_p \beta_{\la_p}({\boldsymbol{\lambda}}) O_p(X)\,.
    \label{eq:traceT}
\end{equation} 
As a result, the $\mathcal{D}_i$'s are no longer protected operators and they get anomalous dimensions. Since all of these operators have the same scaling dimension at ${\boldsymbol{\lambda}}=0$, one needs to solve a mixing problem. In the deformed theory, we denote with the same letter $\mathcal{D}_i$ an arbitrary basis of operators that coincide with the displacement operators in the limit ${\boldsymbol{\lambda}} \to 0$, and with $\overline{\cal D}_i$ the operators with definite boundary scaling dimensions. We have
\be\label{eq:ChangeBasis}
{\cal D}_i = Q({\boldsymbol{\lambda}})_{ij}  \overline{\cal D}_j \,,
\ee
where $Q_{ij}$ is the invertible $n\times n$ mixing matrix.
In terms of $\overline{\cal D}_i$, we have
\begin{equation}\label{eq:DisNorm}
\langle \overline{\mathcal{D}}_i(\vec{x}_1) \overline{\mathcal{D}}_j(\vec{x}_2) \rangle =  \frac{\delta_{ij}}{(\vec{x}_{12}^{\,2})^{\Delta_{i}({\boldsymbol{\lambda}})}}\,, \qquad \Delta_{i}({\boldsymbol{\lambda}}) =d+1 +\gamma_i({\boldsymbol{\lambda}})\,.
\end{equation}
Here $\gamma_i({\boldsymbol{\lambda}})$'s are the anomalous dimensions. The two-point correlators between the $\overline{\calD}_i$ operators and the deformations $O_p$ are fixed by bulk isometries up to a normalization constant, and read
\begin{align}\label{eq:OD2pt}
    \langle O_p(X)\overline{\calD}_i(P)\rangle&=\frac{{C}_{O_p\overline{\calD}_i}({\boldsymbol{\lambda}})}{(-2 P\cdot X)^{\Delta_{_i}({\boldsymbol{\lambda}})}}\,.
\end{align}
While the stress tensor $T$ is no longer traceless, it is still conserved, and there is still only one possible structure in the two-point function with a boundary operator, giving
\begin{align}\label{eq:2ptTDdef}
 \begin{split}
\text{$\lambda_p\neq 0$:}\quad \langle  T_{AB}(X) & \overline{\calD}_i(P) \rangle \\   & = \frac{C_{T\overline{\calD}_i}({\boldsymbol{\lambda}})}{(-2 P\cdot X)^{\Delta_i({\boldsymbol{\lambda}})}}\left(H_{AB}(X,P) - 
 \frac{(\Delta_i({\boldsymbol{\lambda}})-d-1)d}{4(d+1)\Delta_i({\boldsymbol{\lambda}})} G_{AB}(X) \right) \,.
\end{split}
\end{align}
The normalization coefficients ${C}_{O_p\overline{\calD}_i}({\boldsymbol{\lambda}})$ and $C_{T\overline{\calD}_i}({\boldsymbol{\lambda}})$, as well as the scaling dimensions $\Delta_i({\boldsymbol{\lambda}})$, are theory-dependent functions of the couplings $\boldsymbol{\lambda}$.  Imposing \eqref{eq:traceT} and using \eqref{eq:OD2pt}, we get the following $n$ relations between these functions, valid to all orders in the couplings $\lambda_p$:
\begin{empheq}[box=\mybluebox]{align}
\label{eq:AllorderRel}
 -\frac{d(\Delta_i({\boldsymbol{\lambda}})-d-1)}{4\Delta_i({\boldsymbol{\lambda}})}  {C}_{T\overline{\calD}_i}({\boldsymbol{\lambda}}) =   \sum_{p=1}^m \beta_{\lambda_p}({\boldsymbol{\lambda}}) {C}_{O_p\overline{\calD}_i}({\boldsymbol{\lambda}})\,.
\end{empheq}
This is a generalization of (4.12) of \cite{Ciccone:2024guw}, and reduces to it for $m=n=1$. 

Let us now restrict to the case $m=1$ of a single bulk deformation $\lambda$ with associated operator $O$. Expanding \eqref{eq:AllorderRel} in $\lambda$ and keeping only the leading term gives
\begin{equation}\label{eq:PertRel}
  -\frac{d}{4(d+1)}\gamma_i {C}_{T\overline{\cal D}_i}= \beta_0 {C}_{O\overline{\cal D}_i}\,.
  \end{equation}
Here ${C}_{T\overline{\calD}_i}$ and ${C}_{O\overline{\calD}_i}$ denote the constant values approached by the respective functions for $\lambda \to 0$, i.e. their values in the BCFT,\footnote{Note that in BCFT there is a Ward Identity relating ${C}_{O\calD_i}$ to the one-point function of the operator $O$ \cite{Billo:2016cpy}. As a result, as long as there is no symmetry protecting $O$ from having a nonzero one-point function in the BCFT, the perturbative expansion of ${C}_{O\overline{\cal D}_i}(\lambda)$ starts from $\lambda^0$.} while $\beta_0$ and $\gamma_i$ are respectively the coefficient of the first term in the expansion of the $\beta$ function and the anomalous dimension. More specifically, for classically marginal deformations we have $\beta(\lambda)\underset{\lambda\to 0}{\sim}\lambda^2 \beta_0 +\mathcal{O}(\lambda^3)$ and $\Delta_i(\lambda)-d-1\underset{\lambda\to 0}{\sim}\lambda^2 \gamma_i +\mathcal{O}(\lambda^3)$, while for relevant/irrelevant ones both functions start from order $\lambda$. It is interesting to note that, as a consequence of \eqref{eq:AllorderRel}, the expansion of the anomalous dimension starts at subleading order $\lambda^2$ when the bulk coupling is marginal, as opposed to the standard expectation of a leading contribution of order $\lambda$. The scalings mentioned here refer to a generic BCFT deformation, one can readily adapt to special cases by expanding \eqref{eq:AllorderRel}. As we will see below, gauge theories are somewhat special in this respect due to the fact that the coupling appears in the kinetic term of the gauge fields.

The relations \eqref{eq:AllorderRel}-\eqref{eq:PertRel} require knowledge of $\overline{\mathcal{D}}_i$, i.e. of the matrix $Q_{ij}$ defined in \eqref{eq:ChangeBasis}. In practice, however, we do not know a priori what this basis is. 
It is useful to rewrite the expansion \eqref{eq:PertRel} in a basis ${\cal D}_i$
that coincides with the displacement operators for $\lambda \to 0$.
At leading order, the two-point functions between the $\mathcal{D}_i$'s read (here we assume a leading order contribution of order $\lambda$ for definiteness)
\begin{equation}\label{eq:DDExp}
    \langle {\cal D}_i(\vec{x}_1){\cal D}_j(\vec{x}_2)\rangle=\frac{1}{(\vec{x}_{12}^{\,2})^{d+1}}\Big(\delta_{ij}+\lambda M_{ij}-\lambda \Gamma_{ij}\log(\vec{x}_{12}^{\,2}) + \calO(\lambda^2)\Big)\,,
\end{equation}
where $M$ and $\Gamma$ are symmetric $n\times n$ matrices.
Expanding in $\lambda$ \eqref{eq:ChangeBasis} and \eqref{eq:DisNorm}, and matching with \eqref{eq:DDExp}, we get, in matrix notation,
\be
Q(\lambda) = Q_0 - \frac{1}{2} \lambda (M+A)  Q_0+ \frac{1}{2} \lambda  Q_0^T (M-A) + {\cal O}(\lambda^2)\,,
\ee
where $Q_0$ is the $O(n)$ matrix diagonalizing $\Gamma$,
\be
\Gamma = Q_0 \overline\Gamma Q_0^T~,~~\overline\Gamma =\text{diag}\left(\gamma_1,\dots,\gamma_n\right)\,,
\ee
and $A$ is an undetermined (at this order) antisymmetric matrix. As a result, acting with $Q_0$ on both sides of \eqref{eq:PertRel}, we obtain
\begin{empheq}[box=\mybluebox]{align}
    -\frac{d}{4(d+1)}\Gamma_{ij} C_{T{\cal D}_j}= \beta_0 C_{O{\cal D}_i}\,,
    \label{eq:constraintD}
\end{empheq}
which provide $n$ constraints for the $n(n+1)/2$ matrix elements $\Gamma_{ij}$.
The relation \eqref{eq:constraintD} is the generalization of (4.13) of \cite{Ciccone:2024guw} for $n>1$. 

Note that, while for $n=1$ this constraint completely determines the one-loop anomalous dimension in terms of the unperturbed normalization coefficients, it does not for $n>1$. In our main application in this paper we will have $n=2$, meaning that we will get $2$ constraints for $3$ entries of the matrix, and we will need one additional constraint to fix it completely.

\section{Fermionic QCD}
\label{sec:fermions}

In this section we consider fermionic QCD. After a quick review of the pure YM case, we discuss the basic setup for fermions in AdS, boundary conditions, propagators, and the explicit form of the displacement operators. We then proceed to  
solve the one-loop mixing for the latter. 
In section \ref{subsec:CFI_F} we compute it, for both Dirichlet and Neumann bc, by using the results of section \ref{sec:displop}. The anomalous dimensions in the Dirichlet case are then also computed 
by a direct Witten diagram computation in section \ref{subsec:diagramsF}.

\paragraph{Yang-Mills}
We start by recalling the salient features of the theory in the absence of matter, referring the reader to \cite{Ciccone:2024guw} for further details. We adopt the conventions of \cite{Ciccone:2024guw} throughout. 
The action of pure Yang-Mills theory on Euclidean AdS$_{d+1}$, with covariant $\xi$-gauge fixing, reads (omitting the ghost terms) 
\begin{equation}
   S_\text{YM}= \frac{1}{g^2} \int dx\   \mathrm{tr} \bigg[\frac 12 F_{\mu \nu} F^{\mu \nu} +  \frac{1}{\xi} (\nabla_\mu A^\mu)^2\bigg],
   \label{eq:gpAdS1}
\end{equation}
where 
$F_{\mu \nu} = F_{\mu \nu}^a t^a$, with
\begin{equation}
    F_{\mu \nu}^a = \partial_\mu A_\nu^a - \partial_\nu A_\mu^a + f^{a}{}_{bc} A_\mu^b A_\nu^c\,,
    \label{eq:gpAdS2}
\end{equation}
 {${\rm tr} [t^a t^b] = \delta^{ab}/2$}, and $a,b,\dots$ are indices in the adjoint representation of the group. For definiteness, we take the gauge group to be $SU(n_c)$, though our discussion can be straightforwardly generalized.
We are interested in the physics for $d=3$, but keeping $d$ generic is needed for dimensional regularization. 
We use the invariant $u(x,y)$, given by $u = -1 - X \cdot Y $ in embedding coordinates, while in Poincar\'e coordinates $x=(\vec{x},z)$, $y=(\vec{y},w)$ one has $u =[(\vec{x}-\vec{y})^2+(z-w)^2]/(2zw)$.

The bulk-to-bulk gauge propagator is
\begin{equation}
\Pi_{\mu \nu}(x, y)=-g_0(u)\nabla_{\mu}^x\nabla_{\nu}^y u+g_1(u)\nabla_{\mu}^xu\nabla_{\nu}^yu  \,,
\label{eq:propintr}
\end{equation}
where the functions $g_0(u), g_1(u)$ can be determined by solving the bulk equations of motion with the chosen bc. The most natural bc for the gauge field are Dirichlet, defined by 
\begin{equation}
    A_i(x)\underset{z\to0}{\sim} z^{d-2}\, g^2 J_i(\vec{x})\,,
\end{equation}
or Neumann, defined by
\begin{equation}
    A_i(x)\underset{z\to0}{\sim} a_i(\vec{x})\,.
\end{equation}
We denote here with $\mu= i$ ($i = 1, \ldots, d$) the coordinates along the boundary. 
The boundary operator $J_i$ is a conserved current, while $a_i$ is a boundary gauge field. In the Fried-Yennie (FY) gauge, $\xi=d/(d-2)$, the propagators with Dirichlet and Neumann bc take a remarkably simple form. With Dirichlet bc
we have
\begin{align}
\begin{split} g_0^{(\rm D)}(u) & =\frac{ \Gamma \left(\frac{d+1}{2}\right)  }{2 \pi ^{\frac{d+1}{2} } (u (u+2))^{\frac{d-1}{2}}(d-2)}\,,\\ 
     g^{(\rm D)}_1(u)& =\frac{u+1} {u (u+2)} g_0(u)\,,\end{split} 
     \qquad \xi=\frac{d}{d-2}\,,
          \label{subYennie}
\end{align}
while for Neumann bc
\begin{empheq}{align}
\begin{split}
        g_0^{(\rm N)}(u)&=\frac{u (u+2) \, _2F_1\left(1,\frac{d}{2};\frac{d+3}{2};-u (u+2)\right) +(d+1)}{4\pi^{d/2} (d+1) (u+1) \Gamma
   \left(2-\frac{d}{2}\right)}+g_0^{(\rm D)}(u)\,,\\
   g_1^{(\rm N)}(u)&=\frac{u+1}{u(u+2)}\left(g_0^{(\rm N)}(u)-\frac{1}{4\pi^{d/2} (u+1) \Gamma
   \left(2-\frac{d}{2}\right)}\right)\,,
    \end{split}\qquad \xi=\frac{d}{d-2}\,.
          \label{subYennieN}
\end{empheq}
In $d=3$, the propagators in this gauge simplify to
\begin{align}
\begin{split}
g_0^{(\rm D,N)}(u) &= \frac{1}{4\pi^2} \left( \frac{1}{u} \mp \frac{1}{u+2} \right) ,\\
g_1^{(\rm D,N)}(u) &= \frac{1}{8\pi^2} \left( \frac{1}{u^2} \mp \frac{1}{(u+2)^2} \right) ,
\end{split}
\qquad d=3,~~\xi = 3 \,,
\label{eq:prop3}
\end{align}
where the upper sign corresponds to Dirichlet bc and the lower sign to Neumann bc. Thus, in $d=3$ the two boundary conditions differ only by the sign between the two terms in $g_0$ and $g_1$, i.e. we can write both propagators as
\begin{equation}\label{eq:PiGuageProp}
    \Pi^{(\rm D,N)}_{\mu \nu}(x, y) = \Pi^{(1)}_{\mu \nu}(x, y) \; \mp \; \Pi^{(2)}_{\mu \nu}(x, y) \,,
\end{equation}
where $\Pi^{(1)}$ and $\Pi^{(2)}$ follow directly from substituting \eqref{eq:prop3} into \eqref{eq:propintr}. This simple formula, encompassing both propagators, will be useful later to compare the results between Dirichlet and Neumann bc.\footnote{The functions $g_0^{(\text{D,N})}$ and $g_1^{(\text{D,N})}$ are sums or differences of the same function evaluated at the points $u$ and $u+2$. Note that $u+2  = [(\vec{x}-\vec{y})^2+(z+w)^2]/(2zw)$ is obtained from $u$ by sending one point to its reflection with respect to the $z$ coordinate. Intriguingly, the gauge propagator in half flat space with $z\geq 0$ in the FY gauge can be written as in \eqref{eq:propintr}, despite the fact that
the gauge connection is not a well-defined primary operator and hence does not have a well-defined transformation under Weyl rescalings.}

The bulk-to-boundary gauge propagator can be obtained from the bulk-to-bulk propagator by sending one of the two bulk points to the boundary. Using embedding coordinates and up to terms that vanish when projected in physical space, we get, in FY gauge,\footnote{Note that the expression we provide for $K_{AB}^{({\rm N})}(X,P)$ is not manifestly transverse in $P$, that is $P^{B} K_{AB}^{({\rm N})}(X,P) \neq 0 $. This is a manifestation of the fact that the boundary gauge field is not a primary operator.
}
\begin{equation}
\begin{aligned}
    K_{AB}^{(\rm D)}(X,P)& =\frac{  \Gamma(d)}{2\pi^{\frac{d}{2}} (d-2) \Gamma\left(\frac{d}{2}\right)}\frac{(-2P\cdot X)\eta_{AB}+2 P_A X_B}{(-2P\cdot X)^{d}}
    \,,\\
    K_{AB}^{(\rm N)}(X,P)&=\frac{g^2}{2\pi^{\frac {d}2}(d-2)\Gamma(2-\frac d2)}G_{AC}(X)\frac{(d-1)(-2P\cdot X) \eta_{CB}+2P_CX_B}{(-2P\cdot X)^2}\,,
  \end{aligned}
  \label{eq:boundprop}
\end{equation}
and the corresponding expression in physical space can be obtained by projecting
\begin{equation}
 K_{\mu i}(x,\vec{y})=\frac{\de X^A}{\de x^\mu}\frac{\de P^B}{\de y^i}K_{AB}(X,P)\,,
\end{equation}
where in Poincaré coordinates $P^A=(0,\vec{y}^2,y^i)$, $x^\mu=(z,\vec{x})$, and $X^A=\frac1z(1,\vec{x}^2+z^2,x^i)$. 

The bulk Yang-Mills stress tensor reads 
\begin{equation}\label{eq:TstressYM}
    T_{\mu\nu}^{\rm YM}=\frac2{g^2}{\rm tr}\left[F_\mu{}^\rho F_{\nu\rho}-\frac14g_{\mu\nu}F^{\rho\si}F_{\rho\si}\right]\,.
\end{equation}
With Dirichlet bc one finds
\begin{equation}\label{eq:TzzYM}
    T_{zz}^{\rm YM}\underset{z\to0}{\sim} \frac2{g^2}{\rm tr}[(\de_z A_i)(\de_z A_j) g^{ij}-\frac12g_{zz}g^{zz}g^{ij}(\de_z A_i)(\de_z A_j)]=(d-2)^2z^{2(d-2)}{g^2}{\rm tr}[J_i J^i]\,.
\end{equation}
In $d=3$ the bulk free theory is conformal and from \eqref{eq:TzzYM} we obtain that the corresponding displacement operator is
\begin{equation}
    \calD_{\rm YM}=g^2\,{\rm tr}[J_iJ^i]=\frac{g^2}2 J_i^a J^{ai}\,.
    \label{eq:DispYM}
\end{equation}
With Neumann bc one obtains
\begin{equation}\label{eq:DYMN}
    \calD_{\rm YM}=-\frac1{2g^2}{\rm tr}[f_{ij}f^{ij}]=-\frac1{4g^2}f_{ij}^af^{aij}\,,
\end{equation}
with $f_{ij}^a=\de_i a_j^a-\de_j a_i^a +f^a{}_{bc}a^b_ia^c_j$ being the boundary field strength.

\paragraph{Fermions} 
Let us now add fermionic matter. 
We start with the following Euclidean action 
\begin{equation}\label{eq:SymmetricDirac}
S=S_{\rm YM}+ S_{\rm F}\,,\quad S_{\rm F} =-\int dx\,\bar\Psi\left(\frac{1}{2}(\overset{\rightarrow}{\slashed{D}} - \overset{\leftarrow}{\slashed{D}})+M\right)\Psi~.
\end{equation}
$\Psi$ and $\bar{\Psi}$ represent $n_f$ Dirac fermions transforming respectively in the fundamental and anti-fundamental representation 
of the $SU(n_c)$ gauge group. The covariant derivative reads
\begin{equation}
\begin{aligned}
    \overset{\rightarrow}{D}_\mu\Psi_I^\al&= \partial_{\mu}\Psi_I^\al + \frac{\omega^{mn}_{\mu} [\gamma_m, \gamma_n]}{8}\Psi_I^\al+iA_\mu^a (t^a)^\alpha{}_\beta\Psi^\beta_I\,,\\
    \bar\Psi^I_\al\overset{\leftarrow}{D}_\mu&= \partial_{\mu}\bar\Psi^I_\al - \bar\Psi^I_\al\frac{\omega^{mn}_{\mu} [\gamma_m, \gamma_n]}{8}- i A_\mu^a\bar\Psi_\beta^I(t^a)^\beta{}_\al\,.
\end{aligned}
\end{equation}
Here we wrote explicitly the color indices $\alpha,\beta,\dots\,=1,\dots,n_c$ 
and the flavor indices $I,J,\dots=1,\dots,n_f$, that were implicit in \eqref{eq:SymmetricDirac}. We are using $\mu,\nu,\dots$ and $m,n\dots$ to denote respectively the curved-space and flat-space indices. $\Gamma^\mu$ and $\gamma^m$ are the curved-space and flat-space gamma matrices, related to each other by the inverse vielbein:
$\Gamma^\mu = e^\mu_m \gamma^m$. 
In Poincaré coordinates $(z, \vec{x})$, and $e^{\mu}_m = z \delta^{\mu}_m$, so that $\slashed{D}= \Gamma \cdot D = z \gamma^m D_m $. 
The spin connection is given by
\begin{equation}
    \omega^{m n}_{\mu} = \frac{\delta^m_z \delta^n_{\mu} - \delta^n_z \delta^m_{\mu}}{z}\,.
\end{equation}
In order to have a well-defined variational problem, we need to supplement \eqref{eq:SymmetricDirac} with the boundary term
\begin{equation}
S_\partial = \frac12 \int_{z=\epsilon} d^d\vec{x}\,\epsilon^{-d}\bar\Psi^I B_I^{~J}\Psi_J~,
\end{equation}
at a radial cutoff $z=\epsilon \ll 1$, see e.g. \cite{Henningson:1998cd, Arutyunov:1998ve,Henneaux:1998ch}. Taking this term into account, we find that the following boundary conditions
\begin{align}
\begin{split}
 B_I^{~J}\gamma_z
\Psi_J(\epsilon,\vec{x}) & =  \Psi_I(\epsilon,\vec{x})~,\\
\bar{\Psi}^I(\epsilon,\vec{x}) \gamma_z B_I^{~J}& = - \bar{\Psi}^J(\epsilon,\vec{x})~,
\label{eq:fermbc}
\end{split}
\end{align}
make the action stationary on-shell, and therefore they are allowed at least at weak coupling.\footnote{In the case of Dirichlet bc for the gauge field, one can consider more general boundary terms, and more general resulting boundary conditions, that also mix the color indices. This is allowed because the gauge symmetry is a global symmetry at the boundary in this case. If one does that, the gauge symmetry is spontaneously broken. More precisely, there is a Higgs mechanism and the bulk gauge fields gets a mass \cite{Rattazzi:2009ux}.} 
The flavor matrix $B_I^{~J}$ is a free parameter of the boundary term and of the associated boundary condition. Consistency requires that it squares to the identity, i.e. $B_I^{~J} B_J^{~K} = \delta_I^{~K}$. Note that the boundary term, being part of the quadratic action, only plays the role of determining the propagators, via the boundary conditions that it allows. 

Expanding the Dirac equation for small $z$, one finds that the following two modes are allowed for the Dirac field
\begin{align}
\begin{split}
\Psi_I(z,\vec{x}) & \underset{z\to 0}{\sim} z^{\frac d2 + M_I} \widehat{\Psi}_{+\,I}(\vec{x}) + z^{\frac d2 - M_I} \widehat{\Psi}_{-\,I}(\vec{x})~,\\
\bar{\Psi}^I(z,\vec{x}) & \underset{z\to 0}{\sim} z^{\frac d2 + M_I} \widehat{\bar{\Psi}}{}^I_+(\vec{x}) + z^{\frac d2 - M_I} \widehat{\bar{\Psi}}{}^I_-(\vec{x})~,
\end{split}
\end{align}
where the $M_I$'s are the eigenvalues of the mass matrix $M_I^{~J}$, that we assume diagonal in the flavor basis that we are choosing. The coefficient functions $\widehat{\Psi}_{\pm\,I}$ and $\widehat{\bar{\Psi}}{}^I_\pm$ satisfy $\gamma_z \widehat{\Psi}_{\pm\,I} = \mp \widehat{\Psi}_{\pm\,I}$, $\widehat{\bar{\Psi}}{}^I_\pm\gamma_z = \pm \widehat{\bar{\Psi}}{}^I_\pm$. They define boundary operators with scaling dimensions $\Delta_{\pm,I} = \frac d2 \pm M_I$. In order for the boundary condition \eqref{eq:fermbc} to be compatible with AdS isometries, it must be possible to write it in terms of operators with a fixed scaling dimension, i.e. it cannot mix flavor components corresponding to different eigenvalues $M_I$. This means that the matrix $B_I^{~J}$ appearing in the bc must commute with the mass matrix $M_I^{~J}$. As a result they can be diagonalized simultaneously, and we can assume that they are both diagonal in the flavor basis we are choosing. Since it squares to the identity, the only possible eigenvalues of $B_I^{~J}$ are $B_I = \pm 1$. The resulting boundary condition sets $\widehat{\Psi}_{\pm\,I} = 0$ and $\widehat{\bar{\Psi}}{}^I_\pm = 0$, while $\widehat{\Psi}_{\mp\,I}$ and $\widehat{\bar{\Psi}}{}^I_\mp$ are nontrivial boundary operators. In the massless case $M=0$ (or for flavor entries corresponding to zero eigenvalues) we can write more simply the near boundary expansion as
\begin{align}
\begin{split}
\Psi_I(z,\vec{x}) & \underset{z\to 0}{\sim} z^{\frac d2} \widehat{\Psi}_I(\vec{x}) ~,\\
\bar{\Psi}^I(z,\vec{x}) & \underset{z\to 0}{\sim} z^{\frac d2} \widehat{\bar{\Psi}}{}^I(\vec{x})~.
\end{split}
\end{align}
The modes $\widehat{\Psi}_I(\vec{x})$ and $\widehat{\bar{\Psi}}{}^I(\vec{x})$ define boundary operators of scaling dimension $\frac d2$ which satisfy the constraint
\begin{align}
\begin{split}\label{eq:masslessmodes}
 B_I^{~J}\gamma_z
\widehat{\Psi}_J(\vec{x}) & =  \widehat{\Psi}_I(\vec{x})~,\\
\widehat{\bar{\Psi}}{}^I(\vec{x}) \gamma_z B_I^{~J}& = - \widehat{\bar{\Psi}}{}^J(\vec{x})~,
\end{split}
\end{align}
as a consequence of the boundary condition \eqref{eq:fermbc}.

We now proceed to describe the Feynman rules for the fermions that we need for the perturbative calculation. The fermion propagator on  AdS$_{d+1}$ was computed 
e.g. in \cite{Henningson:1998cd,Arutyunov:1998ve,Henneaux:1998ch,Mueck:1998iz,Mueck:1999efk,Kawano:1999au}, here we follow mostly the conventions of \cite{Giombi:2021cnr}. We are interested in the massless case $M=0$, where it reads 
\begin{equation}
\begin{aligned}
S^{\al,J}_{\beta,I}(x,x')&=\langle{\Psi}^\al_I(x)\bar{\Psi}^J_\beta(x')\rangle_\text{AdS}=\delta^\al_{~\beta} S_I^{~J}(x,x')\,,\\ 
S_I^{~J}(x,x')&=\frac{1}{\sqrt{z z'}}\left(\delta_I^{~J}\gamma_m ({x}-x')^m A_1(u)+ B_I^{~J}\gamma_z\gamma_m (\bar{x}-x')^m A_2(u)\right)~,
\label{eq:propfbulk}
\end{aligned}
\end{equation}
where $\bar{x}=(-z,\vec{x})$ is the image point of $x$ with respect to the boundary and $A_1$ and $A_2$ are  
\begin{equation}
A_1(u)=-\frac{\Gamma\left(\frac{d+1}{2}\right)}{2(2\pi)^\frac{d+1}{2}}\frac{1}{u^{\frac{d+1}{2}}}~,\quad A_2(u)= A_1(u+2)~.  
  \label{eq:A&B}
\end{equation}
The bulk-to-boundary propagator is obtained by taking the limit of \eqref{eq:propfbulk}. We have 
\begin{align}
\begin{split}
 \left\langle\widehat{\Psi}^\alpha_I\left(\vec{x}\right) \bar{\Psi}_\beta^J\left(x'\right)\right\rangle & = \delta^\al_{~\beta}(S^{\partial b})_I^{~J}(\vec{x},x')\,,\\
\left\langle{\Psi}^\alpha_I\left({x}\right) \widehat{\bar{\Psi}}{}^{J}_\beta\left(\vec{x}'\right)\right\rangle & =\delta^\al_{~\beta}({S}^{b\partial})_I^{~J}(x,\vec{x}')\,,
\end{split}\label{bulktoboundferm}
\end{align}
where
\begin{align}
\begin{split}
(S^{\partial b})_I^{~J}(\vec{x},x') & =\frac{\delta_I^{~J} + B_I^{~J}\gamma_z}{2} \,\frac{\Gamma\left(\frac{d+1}{2}\right)}{\pi^{\frac{d+1}{2}}}\, \frac{(z')^{\frac d2}\,(\gamma_z z' -\gamma_i (\vec{x}-\vec{x}')^i)}{\left((z')^2+(\vec{x}-\vec{x}')^2\right)^{\frac{d+1}{2}}}\,, \\
 ({S}^{b\partial})_I^{~J}(x,\vec{x}') & =-\frac{\Gamma\left(\frac{d+1}{2}\right)}{\pi^{\frac{d+1}{2}}}\, \frac{z^{\frac d2}\,(\gamma_z z +\gamma_i (\vec{x}-\vec{x}')^i)}{\left((z')^2+(\vec{x}-\vec{x}')^2\right)^{\frac{d+1}{2}}} \,\frac{\delta_I^{~J} - B_I^{~J}\gamma_z}{2} \,.
\end{split}\label{bulktoboundferm2}
\end{align}
Taking also the other point to the boundary, we obtain the boundary-to-boundary propagator 
\begin{align}
\begin{split}
& \left\langle\widehat{\Psi}^\alpha_I\left(\vec{x}_1\right) \widehat{\bar{\Psi}}{}_\beta^J\left(\vec{x}_2\right)\right\rangle = \delta^\al_{~\beta}(S^{\partial\partial})_I^{~J}(\vec{x}_1,\vec{x}_2)\,, \\
&(S^{\partial\partial})_I^{~J}(\vec{x}_1,\vec{x}_2)=-\frac{\delta_I^{~J} + B_I^{~J}\gamma_z}{2} \frac{\Gamma\left(\frac{d+1}{2}\right)}{\pi^{\frac{d+1}{2}}}\frac{\gamma^i \vec{x}_{12,i} }{(\vec{x}_{12}^{\, 2})^{\frac{d+1}{2}}} \,.  
\end{split}
\label{boundtoboundferm}
\end{align}
In this work, we adopt the most symmetric choice for $B_I^{~J}$, namely $B_I^{~J}=\pm \delta_I^{~J}$.  
For a motivation for this choice and a discussion of more general boundary conditions, we refer to subsection ~\ref{subsec:fermbc}.

The bulk fermion stress tensor is given by 
\begin{equation}\label{eq:TstressFer}
    T_{\mu\nu}^{\rm F}=-\frac 14\bar\Psi\left(\Gamma_\mu \overset{\rightarrow}{D}_\nu - \overset{\leftarrow}{D}_\nu \Gamma_\mu - g_{\mu\nu}(\overset{\rightarrow}{\slashed{D}} - \overset{\leftarrow}{\slashed{D}})\right)\Psi + (\mu\leftrightarrow\nu)\,,
\end{equation}
where contraction of the flavor and color indices is left implicit. Assuming Dirichlet bc for the gauge field, one finds
\begin{equation}\label{eq:DispFer}
    \calD_{\rm F}=\frac 12\widehat{\bar\Psi}\overleftrightarrow{\slashed{\de}}\widehat{\Psi}\,,
\end{equation}
where $\overleftrightarrow{\slashed\de}=\ga^i\overset{\rightarrow}{\de}_i-\overset{\leftarrow}{\de}_i\ga^i$. Note that $\calD_F$ is the lightest scalar operator that we can construct with fermions at the boundary which is singlet under the global symmetry and parity even.

\subsection{Broken conformal Ward identities}
\label{subsec:CFI_F}
We now apply the results of section \ref{sec:displop} to an $SU(n_c)$  gauge theory coupled to $n_f$ Dirac fermions in the fundamental representation of the gauge group. 
At zero gauge coupling, the bulk is a direct sum of two free CFTs, the theory of $n_c^2-1$ free gluons and the theory of $n_f n_c$ 
 free massless Dirac fermions.\footnote{Strictly speaking, each of these CFTs is composed
by disjoint free CFTs in AdS, as individual gluons and fermions are decoupled. This further splitting is not relevant here.}
In the notation of section \ref{sec:displop}, we then have $n=2$ and 
\begin{equation}\begin{split}
S_{\text{CFT}_1} & = \left.S_\text{YM}\right\vert_{q}, \qquad S_{\text{CFT}_2} = \left.S_\text{F}\right\vert_{q}, \\
 T^1 & = \left.T^{\rm YM}\right\vert_{q}\,, \qquad  T^2 = \left.T^{\rm F}\right\vert_{q}\,,
\end{split}
\end{equation}
where the subscript $q$ refers to the fact that in the decoupling limit we are keeping only terms that are quadratic in the fields. This case, however, is made somewhat subtle by gauge invariance. Indeed, in order to have a manifestly gauge invariant deformation, it is convenient to work with non-canonically normalized gauge fields, as in \eqref{eq:gpAdS1}. With this choice, the deforming operator is $S_{\text{YM}}$ itself, 
and we do not have a neat separation between the free CFT actions and the interaction as in \eqref{eq:SDeform}. The correct translation to the notation of section \ref{sec:displop} is to take $m=1$, to take the single bulk coupling $\lambda$ to be the gauge coupling $g^2$, and to define the single deforming operator $O$ as the derivative of the interaction with respect to the coupling, namely
\begin{equation}
O = \frac{\partial}{\partial g^2} \left(\frac{1}{2g^2} {\rm tr}\, F^2 \right) = -\frac{1}{2g^4} {\rm tr}\, F^2\,.
\label{eq:O}
\end{equation}

The bulk stress tensor in the interacting theory is 
\begin{equation}
    T_{\mu\nu}=T_{\mu\nu}^{\rm YM}+T_{\mu\nu}^{\rm F}\,,
    \label{eq:T_F}
\end{equation}
and its trace reads, up to curvature terms, 
\begin{equation}
T^\mu_{\phantom{\mu}\mu}(x) =
 \beta_{g^2} O(x) \,,
\label{eq:traceTF}
\end{equation}
where $\beta_{g^2}$ is the gauge coupling beta function. We recall that its one-loop coefficient $\beta_0$ is
\begin{equation}
\beta_{g^2}(g^2) = g^4 \beta_0 + \mathcal{O}(g^6)\,, \qquad \beta_0 = -\frac{1}{(4\pi)^2}\left(\frac{22 n_c}{3}- \frac{4}{3} n_f- \frac{1}{3} n_s\right) \,,
\label{eq:betagF}
\end{equation}
where for later use we also added the contribution of $n_s$ complex scalars, also taken in the fundamental representation of the gauge group.

The displacement operators of the decoupled CFTs at zero coupling are $\calD_{\rm YM}$ and $\calD_{\rm F}$,
where $\calD_{\rm YM}$ is defined in \eqref{eq:DispYM} for Dirichlet bc and in \eqref{eq:DYMN} for Neumann bc, while $\calD_{\rm F}$ can be found in \eqref{eq:DispFer}. When the gauge coupling is turned on,  $\calD_{\rm YM}$ and $\calD_{\rm F}$ are no longer protected and mix.  
We define 
\be \label{eq:dispnorm}
{\cal D}_1 = \frac{1}{c_\text{YM}} {\cal D}_\text{YM} \,, \qquad {\cal D}_2 = \frac{1}{c_\text{F}} {\cal D}_\text{F} \,,
\ee
with
\be \label{eq:cYMcF}
c^2_{\rm YM}=\frac{6}{\pi^4} (n_c^2-1)\,,\qquad c_{\rm F}^2=\frac3{\pi^4}n_f n_c\,,
\ee
so that ${\cal D}_{1,2}$ are normalized as in \eqref{eq:Dis1a}. Note that \eqref{eq:cYMcF} applies for both Dirichlet and Neumann bc. 

For $n=2$, \eqref{eq:constraintD} provides two constraints for the three matrix elements $\Gamma_{ij}$. 
In addition, the entry $\Gamma_{11}$ is straightforward to evaluate at leading order: the two-point function of $\mathcal{D}_1$ is unaffected by the addition 
of matter at order $g^2$, since matter contributions first appear only at order $g^4$. Thus, this entry coincides with the result for the anomalous dimension of $\mathcal{D}_1$ in the pure Yang-Mills theory. From \eqref{eq:AllorderRel} with $n=1$ one gets that a relation analogous to \eqref{eq:constraintD} holds in pure Yang-Mills theory, and it was already used in \cite{Ciccone:2024guw} to determine the anomalous dimension of $\mathcal{D}_1$ in that case. As a result, the whole matrix $\Gamma_{ij}$ can in fact be determined from the knowledge of the tree-level values $C_{O{\cal D}_i}$ and $C_{T{\cal D}_i}$ of the bulk-to-boundary coefficients. 
In our normalization of operators, we have
\be\label{eq:CoDOTexp}
C_{O{\cal D}_i}(g^2) = \frac{C_{O{\cal D}_i}}{g^2}\Big(1 + {\cal O}(g^2)\Big)\,, \qquad C_{T{\cal D}_i}(g^2) = C_{T{\cal D}_i} \Big(1 + {\cal O}(g^2) \Big)\,.
\ee
Computing $C_{O{\cal D}_i}$ and $C_{T{\cal D}_i}$ amounts to a simple Wick contraction, that requires only the terms quadratic in the fields in the expressions   \eqref{eq:O} and \eqref{eq:T_F} for $O$ and $T$. This is a leading order computation and therefore  much easier than the direct computation, which involves Witten diagrams at the next-to-leading order. 
Due to the form of the gauge propagator \eqref{eq:PiGuageProp} in FY gauge, the results for Dirichlet and Neumann bc are related. Using a superscript to distinguish the two cases, we have
\be\label{eq:CTDCOD_DvsN}
C_{T{\cal D}_i}^\text{D} = C_{T{\cal D}_i}^\text{N}\equiv C_{T{\cal D}_i}  \,, \qquad  C_{O{\cal D}_i}^\text{D} = - C_{O{\cal D}_i}^\text{N} \equiv C_{O{\cal D}_i} \,,
\ee
with
\be\begin{split}\label{eq:CTODcoeffD}
    C_{T{\cal D}_1}& =\frac{32}{\sqrt{6}\pi^2}\sqrt{n_c^2-1}\,, \qquad  C_{T{\cal D}_2}= \frac{16}{\sqrt{3}\pi^2}\sqrt{n_f n_c}\,,\\
        C_{O{\cal D}_1}& = -\frac{\sqrt{6}}{\pi^2}\sqrt{n_c^2-1}\,, \qquad \; C_{O{\cal D}_2}= 0\,.
\end{split}\ee
As a consequence, the anomalous dimension matrix in the Dirichlet and Neumann bc case are equal and opposite: 
\be\label{eq:Gamma_DvsN}
\Gamma_{ij}^\text{D} = -\Gamma_{ij}^\text{N} \equiv\Gamma_{ij}  \,.
\ee
The result in pure YM theory \cite{Ciccone:2024guw}  gives
\begin{equation}
   \Gamma_{11} = - \frac{11 n_c}{24\pi^2 }\,.
   \label{eq:Ga11}
\end{equation}
The remaining components are easily determined 
using \eqref{eq:constraintD} and \eqref{eq:CTODcoeffD}. We have  
\be\label{eq:matrixGsc12f}
\Gamma_{12} = \Gamma_{21} = \frac{\sqrt{(n_c^2-1)n_f} }{6\sqrt{2n_c}\pi^2}\,, \qquad \qquad \Gamma_{22}= -\frac{(n_c^2-1)}{6n_c \pi^2}\,.
\ee

Having computed the matrix $\Gamma_{ij}$, we find that its eigenvalues $\gamma_{1,2}$ are
\begin{empheq}[box=\mybluebox]{align}
\begin{split}
\gamma_1&=\frac{-(15 n_c^2-4)-\sqrt{\left(15 n_c^2-4\right)^2 - 32 \left(n_c^2-1\right) n_c\left(\frac{11 n_c}{2}-n_f\right) }}{48 \pi ^2 n_c}~,\\
\gamma_2&=\frac{-(15 n_c^2-4)+\sqrt{\left(15 n_c^2-4\right)^2 - 32 \left(n_c^2-1\right) n_c\left(\frac{11 n_c}{2}-n_f\right) }}{48 \pi ^2 n_c }~.
\end{split}
\label{eq:gammafinalF}
\end{empheq}
The resulting anomalous dimensions of  $\overline{\mathcal{D}}_1$ and $\overline{\mathcal{D}}_2$ with boundary conditions Dirichlet and Neumann are
\begin{empheq}[box=\mybluebox]{align}\label{eq:gammafinalfD}
\gamma_{1}^{\text{D}} & = \gamma_{1}\,,\;\; \qquad \gamma_{2}^{\text{D}}= \gamma_{2}\,, \\
\gamma_{1}^{\text{N}} & = -\gamma_{1}\,, \qquad \gamma_{2}^{\text{N}}= -\gamma_{2}\,.
\label{eq:gammafinalfN}
\end{empheq}
We discuss the physical significance of these results in section \ref{sec:physical}.

\subsection{Direct diagram computation}
\label{subsec:diagramsF}

In this section we perform a direct Witten diagrams calculation of the $2\times 2$ matrix $\Gamma_{ij}$ of anomalous dimensions. This requires computing the matrix of two-point functions of the boundary operators $\mathcal{D}_1$ and $\mathcal{D}_2$ (see \eqref{eq:dispnorm}) at next-to-leading order, and then use \eqref{eq:DDExp} to extract $\Gamma_{ij}$. This independent derivation will allow us to check the results \eqref{eq:gammafinalF} of the previous section. We consider only the case of the Dirichlet bc, for which the calculation is easier than with the Neumann bc. As explained in section \ref{subsec:CFI_F}, $\Gamma_{11}$ can be obtained directly from the pure YM result of \cite{Ciccone:2024guw}. 
For the other components $\Gamma_{12}$ and $\Gamma_{22}$, we need instead to compute new diagrams, see figure \ref{fig:gammafermions}. 
\begin{figure}
\centering\begin{subfigure}[c]{0.14\textwidth}\begin{tikzpicture}
  \begin{feynman}[inline=(a.base)]
  \draw (0,0) circle(1.2);
  \vertex (a) at (0,0) [blob]{};
    \vertex [dot](i1) at(-1.2,0){\small{ }} ;
    \vertex [dot] (b) at (1.2,0){};

    \diagram* {
      (i1)-- [fermion,quarter left](b)--[fermion](a)--[fermion](i1)
    };
  \end{feynman}
\end{tikzpicture} 
        
        \caption{}
        \end{subfigure}  \qquad
\begin{subfigure}[c]{0.14\textwidth}
                    \begin{tikzpicture}
  \begin{feynman}[inline=(a.base)]
  \draw (0,0) circle(1.2);
    \vertex [dot](i1) at (-1.2,0){\small{ }} ;
    \vertex[dot](i2) at (0,0.6){\small{ }} ;
    \vertex [dot] (b) at (1.2,0){};
    \vertex [dot] (o2) at (0,-0.6){} ;

    \diagram* {
      (i1)-- [fermion] (i2)-- [fermion] (b)-- [fermion] (o2) -- [fermion] (i1),(o2)--[gluon](i2)
    };
  \end{feynman}
\end{tikzpicture}\caption{} \end{subfigure}\qquad 
 \begin{subfigure}[c]{0.14\textwidth} \begin{tikzpicture}
  \begin{feynman}[inline=(a.base)]
  \draw (0,0) circle(1.2);
    \vertex [dot](i1) at (-1.2,0){\small{ }} ;
    \vertex[dot](i2) at (0,0.6){\small{ }} ;
    \vertex [dot] (b) at (1.2,0){};
    \vertex [dot] (o2) at (0,-0.6){} ;

    \diagram* {
      (i1)-- [gluon] (i2)-- [fermion] (b)-- [fermion] (o2) -- [gluon] (i1),(o2)--[fermion](i2)
    };
  \end{feynman}
\end{tikzpicture}\caption{} \end{subfigure}
        \caption{Witten diagrams that contribute to the two point function of $\calD_{\rm F}$ (a, b) and to the mixing two point function between $\calD_{\rm YM}$ and $\calD_{\rm F}$ (c) at the next-to-leading order.}
        \label{fig:gammafermions}
        \end{figure}

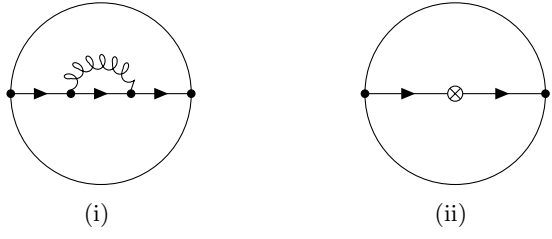
\begin{figure}[t!]
  \centering 
    \renewcommand{\thesubfigure}{\roman{subfigure}}
            \begin{subfigure}[c]{0.15\textwidth}
\begin{minipage}{4cm}

                    \begin{tikzpicture}
  \begin{feynman}[inline=(a.base)]
  \draw (0,0) circle (1.2);
  \vertex (a) at (-0.4,0) [dot]{\small{ }};
    \vertex (i1) at (-1.2,0)[dot]{\small{ }} ;
    \vertex (b) at (0.4,0)[dot]{\small{ }};
    \vertex  (o2) at (1.2,0)[dot] {\small{  }};

    \diagram* {
      (i1) -- [fermion] (a),
      (a) -- [fermion] (b), (a)-- [gluon
, half left]  (b),
      (b) -- [fermion] (o2),
    };
  \end{feynman}
\end{tikzpicture} 
        \end{minipage}
        \caption{}
        \end{subfigure}\qquad\qquad\qquad
            \begin{subfigure}[c]{0.15\textwidth}
                \begin{tikzpicture}
  \begin{feynman}[inline=(a.base)]
  \draw (0,0) circle(1.2);
  \vertex (a) at (0,0) [crossed dot]{};
    \vertex [dot](i1) at(-1.2,0){\small{ }} ;
    \vertex [dot] (b) at (1.2,0){};

    \diagram* {
      (i1)--[fermion](a)--[fermion](b)
    };
  \end{feynman}
\end{tikzpicture} 
     \caption{}
    \end{subfigure}
\caption{One-loop corrections to the fermion two-point function. The counterterm contribution (ii) does not give rise to logarithmic terms and hence does not contribute to the anomalous dimension.
}
\label{fig:twopointfermion}
\end{figure}

We start from the one-loop correction to the fermion two-point function, entering in the diagram (a) in figure \ref{fig:gammafermions}. There are two contributions in this case, see figure \ref{fig:twopointfermion}, but only one gives a non-vanishing contribution to $\Gamma$. Indeed, the counterterm, which is associated with the wavefunction renormalization of the fermionic field, simply rescales the overall normalization of the two-point function. As a result it does not affect the coefficient of the logarithmic term in \eqref{eq:DDExp}. This is expected because otherwise the final result would depend on the renormalization scheme, which is incompatible with the requirement that physical quantities have to be scheme-independent.\footnote{While the scaling dimension is an observable, its coefficients in the expansion in the scheme-dependent quantity $g^2$ can be themselves scheme-dependent. However this type of scheme dependence does not affect the leading order in $g^2$.} The remaining diagram reads
\begin{equation}
\langle{\widehat\Psi}_I^{\alpha_1}(\vec{x}_1) \widehat{\bar\Psi}{}^J_{\alpha_2}(\vec{x}_2)\rangle^{(1)}
 =g^2C_F\di^{\al_1}_{~\al_2} \delta_I^{~J}\!\intads\! dx\, dy\, S^{\partial b}(\vec{x}_2, x) \Gamma^\mu S(x,y)\Gamma^\nu S^{b \partial}(y,\vec{x}_1)\Pi_{\mu \nu}(x,y)\,,
\label{eq:twoPferm}
\end{equation}
where $C_F=(n_c^2-1)/(2 n_c)$ is the quadratic Casimir of the fundamental representation.
The term \eqref{eq:twoPferm} results into a linear combination of integrals with a given number of open Lorentz indices contracted with the product of as many Dirac matrices. Such integrals can in principle be reduced using Lorentz invariance, while the products of Dirac matrices can be rearranged by using the Clifford algebra. Still, as the number of open indices can be considerably high, this procedure may be long and complicated. To make it easier one can use the fact that the result of the integral has to be proportional to the two-point function of the fermion boundary operator. Conformal invariance and the condition \eqref{eq:masslessmodes} force this two-point function to have the form 
\begin{equation}
\langle\widehat{\Psi}_I^{\alpha_1}(\vec{x}_1) \widehat{\bar{\Psi}}{}^J_{\alpha_2}(\vec{x}_2)\rangle= -  \delta_{~\alpha_2}^{\alpha_1} \frac{\delta_I^{~J} + \gamma_z B_I^{~J}}{2}\frac{\Gamma\left(\frac{d+1}{2}\right)}{\pi^{\frac{d+1}{2}}}  \frac{\gamma^i \vec{x}_{12,i} }{(\vec{x}_{12}^{\, 2})^{\Delta_\psi+\frac 12}}\Big(1+{\cal O}(g^2) \Big)  \,,
\label{boundtoboundfermInt}
\end{equation}
where $\Delta_\psi$ is the scaling dimension of $\widehat \Psi$. We parametrize it for small $g^2$ as
\be
\Delta_\psi = \frac{d}2- g^2 C_F C_\psi + {\cal O}(g^4)\,,
\ee
with
$C_\psi$ a constant to be determined. In other words, we expect
\begin{equation}\label{eq:ferm2ptCont1}
\begin{aligned}
\langle \widehat{\Psi}_I^{\alpha_1}(\vec{x}_1) \widehat{\bar\Psi}{}^J_{\alpha_2}(\vec{x}_2)\rangle^{(1)}\Big|_{\log} &
 =g^2 C_F\di^{\al_1}_{~\al_2} \delta^{~J}_I \ C_\psi\ S^{\partial\partial}(\vec{x}_1,\vec{x}_2)\,,
\end{aligned}
\end{equation}
where $S^{\partial\partial}$ is the free boundary-to-boundary propagator \eqref{boundtoboundferm}.
To compute the value of $C_\psi$, one can multiply \eqref{eq:ferm2ptCont1} by the curved-space gamma matrix $\Gamma_\kappa$,  trace over spinor indices, and substitute \eqref{eq:twoPferm} on the left-hand side. This gives
\begin{equation}
C_\psi\Tr\left[ S^{\partial\partial}(\vec{x}_1,\vec{x}_2)\Gamma_\kappa\right]=\left.\Tr{}\left[\intads dx\, dy\, S^{\partial b}(\vec{x}_2, x) \Gamma^\mu S(x,y)\Gamma^\nu S^{b \partial}(y,\vec{x}_1)\Pi_{\mu \nu}(x,y)\Gamma_\kappa\right]\right|_\mathrm{log}\,.
\end{equation}
We compute the traces of Dirac matrices using the \texttt{Mathematica}  package \texttt{FeynCalc} \cite{Mertig:1990an,Shtabovenko:2016sxi,Shtabovenko:2020gxv}.
This leaves only integrals with a single open Lorentz index to be evaluated, much easier to compute. Indeed, they can be straightforwardly reduced, by exploiting Lorentz invariance, to a linear combination of scalar integrals of the form
\begin{equation}
   \mathcal{K}=\int dX \, dY (-2P_1\cdot X)^{-\Delta_1} (-2P_1\cdot Y)^{-\Delta_3} (-2P_2\cdot X)^{-\Delta_2} (-2P_2\cdot Y)^{-\Delta_4} f\left(u\left(X, Y\right)\right)\,.
   \label{eq:ref_int}
\end{equation}
As explained in section 4.2.2 of \cite{Ciccone:2024guw}, these integrals can be written as a double infinite sum of ratios of gamma functions, which in this case can be computed analytically. We then get 
\begin{equation}
    C_\psi=\frac{3}{16\pi^2}\,.
\end{equation}
We can now embed this result in the diagram (a) of figure \ref{fig:gammafermions}. Recalling that the operator ${\cal D}_\text{F}$ in \eqref{eq:DispFer} has derivatives acting either on the incoming or on the outgoing boudary fermion, and that both the upper and lower propagators get one-loop corrections, we get 
\begin{equation}\label{eq:Fermpanela}
\begin{aligned}
\left.\langle \calD_{\rm F}(\vec{x}_1)\calD_{\rm F}(\vec{x}_2)\rangle^{(1)}\right|^{(a)}_{\rm log} & = 
\left.-\frac12\Tr\left[\langle\widehat\Psi_I^{\alpha_1}(\vec{x}_1)\widehat{\bar\Psi}{}^J_{\alpha_2}(\vec{x}_2)\rangle^{(0)}\overleftrightarrow{\slashed{\partial}}_{x_2}\langle\widehat\Psi_J^{\alpha_2}(\vec{x}_2)\widehat{\bar\Psi}{}^I_{\alpha_1}(\vec{x}_1)\rangle^{(1)}\overleftrightarrow{\slashed{\partial}}_{x_1}\right]\right|_{\rm{log }}\\
&=- g^2 n_f \left(n_c^2-1\right) \frac{9}{16 \pi^6 (\vec{x}^{\, 2}_{12})^4}\,.
\end{aligned}
\end{equation}

Next, we consider the diagram (b) in figure \ref{fig:gammafermions}, which amounts to 
\begin{align}      \label{eq:DfermA}
    \left.\langle \calD_{\rm F}(\vec{x}_1)\calD_{\rm F}(\vec{x}_2)\rangle^{(1)}\right|^{(b)}= &\frac{g^2}{8}  n_f (n_c^2-1) \intads dx\ dy\ \Pi_{\mu\nu}(x,y)\\
     &\Tr\left[\Gamma^\mu\left(S^{b \partial}\left(x, \vec{x}_2\right) \overleftrightarrow{\slashed{\partial}}_{{x}_2} S^{\partial b}\left(\vec{x}_2, y\right)\right)\Gamma^\nu\left(S^{b \partial}\left(y, \vec{x}_1\right) \overleftrightarrow{\slashed{\partial}}_{{x}_1} S^{\partial b}\left(\vec{x}_1, x\right)\right)\right] \,. \nn
    \end{align}
  Again, we compute the traces of Dirac matrices with \texttt{FeynCalc} and reduce the integral to a linear combination of scalar integrals of the form \eqref{eq:ref_int}. For this diagram, as well as for the subsequent ones, the resulting double infinite sum is more intricate than in the first diagram and cannot be computed in closed form. Nevertheless, it is possible to reconstruct the analytic expression of the result by observing that it must take the form of a rational function divided by an appropriate power of $\pi$.\footnote{To compute the overall power of $\pi$ we assign a factor of $\pi^{-2}$ to each propagator in the diagram and a factor $\pi^2$ to each point in the bulk over which we integrate.} With this procedure, we find
\begin{equation}
    \left.\langle \calD_{\rm F}(\vec{x}_1)\calD_{\rm F}(\vec{x}_2)\rangle^{(1)}\right|_{\rm log}^{(b)}\simeq g^2  n_f (n_c^2-1) \frac{17}{16 \pi^6 (\vec{x}^{\,2}_{12})^4} \,,
\end{equation}
where here and throughout the paper $\simeq$ denotes the analytic expression extracted from our numerical evaluation. The right-hand side matches the numerical result at least at the sixth significant digit, supporting our identification.

Let us now focus on the mixing diagram (c) in figure \ref{fig:gammafermions}, 
\begin{equation}
\begin{aligned}
    \left.\langle \calD_{\rm YM}(\vec{x}_1)\calD_{\rm F}(\vec{x}_2)\rangle^{(1)}\right|^{(c)}=&\frac{g^2}{4}  n_f (n_c^2-1)\! \intads \!dx\ dy\  K^{i}_{\ \mu}(\vec{x}_1,x)K_{i \nu}(\vec{x}_1,y)\\&\Tr\left[\Gamma^\mu S\left(x, y\right)\Gamma^\nu S^{b \partial}\left(y, \vec{x}_2\right) \overleftrightarrow{\slashed{\partial}}_{{x}_2} S^{ \partial b}\left(\vec{x}_2, x\right)\right] \,.
    \end{aligned}
    \label{eq:DfermA}
\end{equation}
Proceeding as before, we get
\begin{equation}
    \left.\langle \calD_{\rm YM}(\vec{x}_1)\calD_{\rm F}(\vec{x}_2)\rangle^{(1)}\right|_{\rm log}^{(c)}\simeq-g^2  n_f (n_c^2-1) \frac{1}{ 2\pi^6 
    (\vec{x}^{\, 2}_{12})^4} \,.
\end{equation}

Putting all the results together and rescaling the operators as in \eqref{eq:dispnorm}, we get
\begin{equation}
\begin{split}
\Gamma_{22}& =-g^{-2}(\vec{x}^{\, 2}_{12})^4\left.\langle \calD_{2}(\vec{x}_1)\calD_{2}(\vec{x}_2)\rangle^{(1)}\right|_{\rm log}\simeq -\frac{(n_c^2-1)}{6n_c \pi^2} \,, \\
\Gamma_{12}&=\Gamma_{21}=-g^{-2}(\vec{x}^{\, 2}_{12})^4\left.\langle \calD_{1}(\vec{x}_1)\calD_{2}(\vec{x}_2)\rangle^{(1)}\right|_{\rm log}\simeq \frac{\sqrt{(n_c^2-1)n_f} }{6\sqrt{2n_c}\pi^2}\,,
\end{split}
\end{equation}
which reproduces \eqref{eq:matrixGsc12f} obtained using broken conformal Ward identities, providing a non-trivial check of the result.

\section{Scalar QCD}
\label{sec:scalars}

In this section we study scalar QCD. The analysis follows the same lines of section \ref{sec:fermions}, so we will be briefer.
After a quick review of the basic setup for scalars in AdS, we compute the explicit form of the displacement operators and  
solve their mixing at one-loop.
Like in the fermionic case, we 
perform the latter by using the results of section \ref{sec:displop} for both Dirichlet and Neumann bc in section \ref{subsec:CFI_S},  
as well as by a direct Witten diagram computation for Dirichlet bc in section \ref{subsec:diagramsS}.

The action of $n_s$ massive scalar fields coupled to a non-abelian gauge theory reads
\begin{equation}
\label{eq:actionSC}
S = S_{\rm YM} + S_{\rm SC}\,,\qquad 
S_{\rm SC} = \int dx\,\left( D_\mu \phi^*{}^I_\alpha  D_\mu \phi^\alpha_I - 
m^2 \phi^*{}^I_\alpha \phi^\alpha_I \right)\,, 
\end{equation}
with $I = 1, \ldots, n_s$ and $\alpha = 1, \ldots, n_c$ 
denoting the flavor and color indices, respectively. The covariant derivative $D_\mu$ is defined as
\begin{align}
D_\mu \phi_I^\alpha = \nabla_\mu \phi_I^\alpha - i A_\mu^a \left( t^a \right)^{\alpha}_{~\beta} \phi_I^\beta\,, \qquad
D_\mu \phi^*{}^I_\alpha = \nabla_\mu \phi^*{}^I_\alpha + i A_\mu^a  \phi^*{}^I_\beta\left( t^a\right)^{\beta}_{~\alpha}\,.
\end{align}
For simplicity we neglect the quartic self-interacting scalar potential. Such interaction are generated at one-loop level, but 
this effect is subleading in our considerations.

The bulk-to-bulk propagator of a scalar field of mass $m^2 = \Delta(\Delta -d)$ in AdS$_{d+1}$, with boundary condition $\phi\underset{z\to 0}{\sim} z^\Delta \widehat\phi$, reads
\begin{equation}
G(u) = \frac{\Gamma(\Delta)}{2 \pi^{\frac{d-1}{2}} \Gamma\left( \Delta + \frac{2 - d}{2} \right) (2u)^\Delta}
 {}_2F_1\left( \Delta, \Delta - \frac{d+1}{2}, 2\Delta - d + 1, -\frac{2}{u} \right)\,.
\label{eq:propsbulk}
\end{equation}
The bulk-to-boundary propagator is
\begin{equation}
K(x,\vec{y}) = C_0(\Delta) \left(\frac{z}{z^2+(\vec{x}-\vec{y})^2}\right)^\Delta\,, \quad
C_0(\Delta) = \frac{\Gamma(\Delta)}{2 \pi^{\frac{d}{2}} \Gamma\left( 1 - \frac{d}{2} + \Delta \right)}\,.
\label{eq:propsbound}
\end{equation}
The parameter $\Delta$ is the scaling dimension of the boundary operator $\widehat{\phi}$ in the decoupling limit $g^2=0$. We will consider the case of conformally coupled scalars, $m^2=(d^2-1)/4$, so that $g^2$ is the only coupling breaking bulk conformal symmetry. There are two possible boundary conditions for a conformally coupled scalar field, corresponding to $\Delta = (d \pm 1)/2$
. We focus on the case $\Delta = (d + 1)/2$. Like in the fermionic case, we impose this condition on all the (color and) flavor components to maximize the boundary symmetry.

The bulk stress tensor for the action \eqref{eq:actionSC} of conformally coupled scalars reads
\begin{equation}\label{eq:TstressScal}
\begin{aligned}
    T_{\mu\nu}^{\rm SC}&={\rm tr}\left[D_\mu\phi^*D_\nu\phi + D_\nu\phi^*D_\mu\phi-g_{\mu\nu}(D_\la \phi^*D^\la \phi)\right]\\
    &\qquad+\frac{d-1}{2d}\left(\frac{d(d-1)}2g_{\mu\nu}+g_{\mu\nu}\nabla^2-\nabla_\mu \nabla_\nu\right){\rm tr}\,[\phi^*\phi]\,,
\end{aligned}
\end{equation}
where the trace is over color and flavor indices. Using the free theory boundary OPE $\phi\underset{z\to 0}{\sim} z^{\frac{d+1}2} \widehat\phi$ we get 
\begin{equation}\label{eq:DispScal}
    \calD_{\rm SC}=\,{\rm tr}\left[\widehat\phi^*\widehat\phi\right]\,.
\end{equation} 
Note that this is the lightest scalar operator that we can construct from boundary scalars which is a singlet under the boundary symmetry. 

\subsection{Broken conformal Ward identities}
\label{subsec:CFI_S}
We can repeat verbatim the analysis of section \ref{subsec:CFI_F} for the case of scalars. The only difference lies in the definition of the operator $\calD_2$, which now we take to be
\be
\calD_2=\frac1{c_{\rm SC}}\calD_{\rm SC}\,,\qquad c^2_{\rm SC}=\frac1{\pi^4}n_s n_c\,,
\label{eq:dispnormS}
\ee
where $\calD_{\rm SC}$ has been defined in \eqref{eq:DispScal}. As a consequence, the corresponding quantities in \eqref{eq:CTODcoeffD} are now computed to be
\be
C_{T\calD_2}=-\frac{16}{3\pi^2}\sqrt{n_s n_c}\,,\qquad C_{O\calD_2}=0\,.
\ee
In particular, the relations \eqref{eq:CTDCOD_DvsN}, \eqref{eq:Gamma_DvsN}, and \eqref{eq:Ga11} still hold. Using \eqref{eq:constraintD}, we determine the missing entries of $\Gamma_{ij}$,
\be\label{eq:matrixGsc12}
\Gamma_{12} = \Gamma_{21} = \frac{\sqrt{(n_c^2-1)n_s}}{8\sqrt{6n_c}\pi^2}\,, \qquad \qquad \Gamma_{22}= -\frac{(n_c^2-1)}{8n_c \pi^2}\,,
\ee
and then its eigenvalues,
\begin{empheq}[box=\mybluebox]{align}
\begin{split}
\gamma_1&=\frac{-(14 n_c^2-3)-\sqrt{\left(14 n_c^2-3\right)^2 - 6 \left(n_c^2-1\right) n_c(22 n_c - n_s)}}{48 \pi ^2 n_c}\,,\\
\gamma_2&=\frac{-(14 n_c^2-3)+\sqrt{\left(14 n_c^2-3\right)^2 - 6 \left(n_c^2-1\right) n_c(22 n_c - n_s) }}{48 \pi ^2 n_c}\,.
\end{split}
\label{eq:gammafinalS}
\end{empheq}
Once again, $\gamma_i^{\rm D}$ and $\gamma_i^{\rm N}$ are related to the above by \eqref{eq:gammafinalfD} and \eqref{eq:gammafinalfN}, respectively.

\subsection{Direct diagram computation}
\label{subsec:diagramsS}

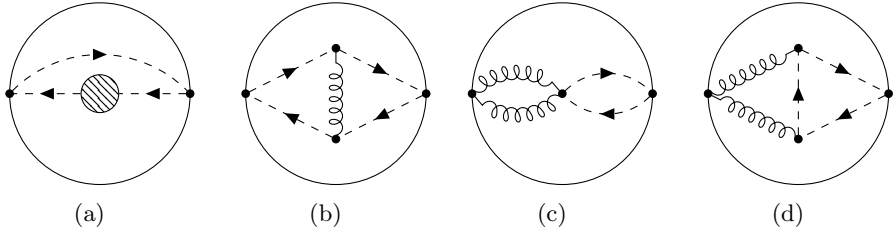
\begin{figure}[t!]
\centering\begin{subfigure}[c]{0.14\textwidth}\begin{tikzpicture}
  \begin{feynman}[inline=(a.base)]
  \draw (0,0) circle(1.2);
  \vertex (a) at (0,0) [blob]{};
    \vertex [dot](i1) at(-1.2,0){\small{ }} ;
    \vertex [dot] (b) at (1.2,0){};

    \diagram* {
      (i1)-- [charged scalar,quarter left](b)--[charged scalar](a)--[charged scalar](i1)
    };
  \end{feynman}
\end{tikzpicture} 
        
        \caption{}
        \end{subfigure}  \qquad
\begin{subfigure}[c]{0.14\textwidth}
                    \begin{tikzpicture}
  \begin{feynman}[inline=(a.base)]
  \draw (0,0) circle(1.2);
    \vertex [dot](i1) at (-1.2,0){\small{ }} ;
    \vertex[dot](i2) at (0,0.6){\small{ }} ;
    \vertex [dot] (b) at (1.2,0){};
    \vertex [dot] (o2) at (0,-0.6){} ;

    \diagram* {
      (i1)-- [charged scalar] (i2)-- [charged scalar] (b)-- [charged scalar] (o2) -- [charged scalar] (i1),(o2)--[gluon](i2)
    };
  \end{feynman}
\end{tikzpicture}\caption{} \end{subfigure}\qquad 
        \begin{subfigure}[c]{0.14\textwidth} \begin{tikzpicture}
  \begin{feynman}[inline=(a.base)]
  \draw (0,0) circle(1.2);
    \vertex [dot](i1) at (-1.2,0){\small{ }} ;
    \vertex[dot](i2) at (0,0){\small{ }} ;
    \vertex [dot] (o2) at (1.2,0){} ;

    \diagram* {
      (i1)-- [gluon, quarter left] (i2)-- [charged scalar, quarter left] (o2)-- [charged scalar, quarter left] (i2) -- [gluon, quarter left] (i1)
    };
  \end{feynman}
\end{tikzpicture}\caption{} \end{subfigure}
  \qquad  \begin{subfigure}[c]{0.14\textwidth} \begin{tikzpicture}
  \begin{feynman}[inline=(a.base)]
  \draw (0,0) circle(1.2);
    \vertex [dot](i1) at (-1.2,0){\small{ }} ;
    \vertex[dot](i2) at (0,0.6){\small{ }} ;
    \vertex [dot] (b) at (1.2,0){};
    \vertex [dot] (o2) at (0,-0.6){} ;

    \diagram* {
      (i1)-- [gluon] (i2)-- [charged scalar] (b)-- [charged scalar] (o2) -- [gluon] (i1),(o2)--[charged scalar](i2)
    };
  \end{feynman}
\end{tikzpicture}\caption{} \end{subfigure}
        \caption{Witten diagrams that contribute to the two point function of $\calD_{\rm SC}$ (a, b) and to the mixing two point function between $\calD_{\rm YM}$ and $\calD_{\rm SC}$ (c, d) at the next-to-leading order. }
        \label{fig:gammascalars}
        \end{figure}

For the explicit Witten diagram computation of $\Gamma_{12}$ and $\Gamma_{22}$, we consider again only the case of Dirichlet bc.
Following \eqref{eq:DDExp}, we need to compute the logarithmic term of the two-point function of ${\calD_{\rm SC}}$ and of the mixing two-point function between ${\calD_{\rm SC}}$ and ${\calD_{\rm YM}}$. The relevant diagrams are shown in figure  \ref{fig:gammascalars}.
Let us begin with the one-loop correction to the two-point function of the scalar field, entering in the diagram (a) in figure \ref{fig:gammascalars}.

In a generic gauge, there are three contributions to this diagram, see figure \ref{fig:twopointscalar}: two from the triple and quartic scalar-gauge interactions, panels (i) and (ii), and one from the counterterm, panel (iii) in figure \ref{fig:twopointscalar}. When working in the FY gauge, the gauge-field tadpole vanishes in dimensional regularization. As a result, the quartic interaction in (ii) does not contribute and can be neglected. 
For conformally coupled scalars, the only required counterterm in (iii) is associated with the wavefunction renormalization of the scalar field \cite{Brown:1980qq,Hathrell:1981zb,Jack:1985wd}. 
As in the fermionic case, this counterterm simply rescales the overall normalization of the two-point function and does not affect the coefficient of the logarithmic term. 
We are therefore left only with the computation of the diagram (i) in figure \ref{fig:twopointscalar}.
This evaluates to
\begin{align}\begin{split}
&\langle\phi_I^{\alpha_1}(\vec{x}_1) \phi^*{}^J_{\alpha_2}(\vec{x}_2)\rangle^{(1)}
 =-g^2C_F\di^{\al_1}_{~\al_2} \delta_{I}^{~J}\intads dx\, dy\, \Big(K(\vec{x}_1, x) \nabla_\mu^x G(x, y) \nabla_\nu^y K(y, \vec{x}_2)\\
&-K(\vec{x}_1, x) \nabla_\mu^x \nabla_\nu^y G(x, y) K(y, \vec{x}_2) \ -\nabla_\mu^x K\left(\vec{x}_1, x\right) G(x, y) \nabla_\nu^y K\left(y, \vec{x}_2\right)\\& +\nabla_\mu^x K\left(\vec{x}_1, x\right) \nabla_\nu^yG(x, y) K\left(y, \vec{x}_2\right)\Big)\Pi^{\mu \nu}(x,y) \,.
\end{split}
\end{align}
For reasons that will become clear soon, it is convenient to remove derivatives from the external scalar propagator by performing integration by parts.\footnote{We verified that the boundary terms vanish, which justifies moving derivatives as if the space had no boundary.} This leads to 
\begin{equation}\begin{aligned}
&\langle\phi_I^{\alpha_1}(\vec{x}_1) \phi^*{}^J_{\alpha_2}(\vec{x}_2)\rangle^{(1)}
 =g^2C_F\di^{\al_1}_{~\alpha_2} \delta_{I}^{~J}\intads dx\, dy\,  K(\vec{x}_1, x)K(y,\vec{x}_2)\Big(2 \nabla_\mu^x  \Pi^{\mu \nu}(x,y) \nabla_\nu^y G(x,y)\\
&+2 \nabla_\mu^x G(x,y)\nabla_\nu^y \Pi^{\mu \nu}(x,y) +4 \nabla_\mu^x \nabla_\nu^y G(x, y)\Pi^{\mu \nu}(x,y)+\nabla_\mu^x \nabla_\nu^y \Pi^{\mu \nu}(x, y) G(x,y)\Big) ~.
\end{aligned}
\label{eq:twopointscalar1L}
\end{equation}
Replacing the expression of the propagators, we find that the integrand vanishes, so no complicated integral has to be computed. However, the presence of double derivatives acting on propagators may generate contact terms which cannot be neglected \cite{DeCesare:2022obt}. Contact terms due to the action of double derivatives on scalar propagators, the next to last term in \eqref{eq:twopointscalar1L}, generate vector tadpoles which, as explained before, vanish in FY gauge. Contact terms due to the action of double derivatives on gauge propagators, i.e. the last term in \eqref{eq:twopointscalar1L}, are instead non-vanishing. 
From the e.o.m. for the vector propagator in a generic $\xi$-gauge we get
\begin{figure}[t!]
  \centering 
  \renewcommand{\thesubfigure}{\roman{subfigure}}
            \begin{subfigure}[c]{0.15\textwidth}
\begin{minipage}{4cm}

                    \begin{tikzpicture}
  \begin{feynman}[inline=(a.base)]
  \draw (0,0) circle (1.2);
  \vertex (a) at (-0.4,0) [dot]{\small{ }};
    \vertex (i1) at (-1.2,0)[dot]{\small{ }} ;
    \vertex (b) at (0.4,0)[dot]{\small{ }};
    \vertex  (o2) at (1.2,0)[dot] {\small{  }};

    \diagram* {
      (i1) -- [charged scalar] (a),
      (a) -- [charged scalar] (b), (a)-- [gluon
, half left]  (b),
      (b) -- [charged scalar] (o2),
    };
  \end{feynman}
\end{tikzpicture} 
        \end{minipage}
        \caption{}
        \end{subfigure}\qquad \begin{subfigure}[c]{0.15\textwidth}
  \begin{minipage}{4cm}
\begin{tikzpicture}
  \begin{feynman}[inline=(d.base)]
  \vertex (a) at (0,0)[dot] {\small{}};
   \vertex (i1) at (-1.2,0)[dot]{\small{ }} ;
    \vertex (b) [dot] at (1.2,0){\small};
  \vertex (c) at (0,2/3);
    \draw (0,0) circle (1.2);

    \diagram* {
      (i1) -- [charged scalar] (a),
      (a) -- [gluon, half left] (c),
      (c) -- [gluon, half left] (a),
      (a) -- [charged scalar] (b),
    };
  \end{feynman}
\end{tikzpicture}  
\end{minipage}
        \caption{}
    \end{subfigure}\qquad
            \begin{subfigure}[c]{0.15\textwidth}
                \begin{tikzpicture}
  \begin{feynman}[inline=(a.base)]
  \draw (0,0) circle(1.2);
  \vertex (a) at (0,0) [crossed dot]{};
    \vertex [dot](i1) at(-1.2,0){\small{ }} ;
    \vertex [dot] (b) at (1.2,0){};

    \diagram* {
      (i1)--[charged scalar](a)--[charged scalar](b)
    };
  \end{feynman}
\end{tikzpicture} 
     \caption{}
    \end{subfigure}
\caption{One-loop corrections to the scalar two-point function. The quartic diagram (ii) vanishes in the FY gauge, while the counterterm (iii) contribution does not give rise to logarithmic terms.}
\label{fig:twopointscalar}
\end{figure}
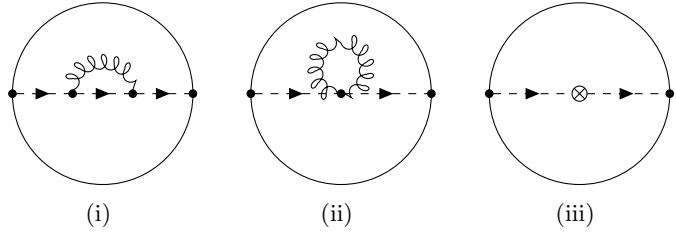
\begin{equation}
    \nabla_\mu^x \nabla_\nu^y \Pi^{\mu \nu}(x, y)=\xi\  \delta(x-y)\,+\dots, 
\end{equation}
where the dots denote terms which are regular at coincident points. 
This gives, for the FY gauge in $d=3$,
\begin{equation}\begin{aligned}
\left.\langle\phi_I^{\alpha_1}(\vec{x}_1) \phi^*{}^J_{\alpha_2}(\vec{x}_2)\rangle^{(1)}\right|_{\log} &
=\left.3g^2C_F\di^{\al_1}_{~\alpha_2} \delta_{I}^{~J}\intads dx\, dy\,  K(\vec{x}_1, x)K(y,\vec{x}_2)G(x,y)\delta(x,y) \right|_{\log }\\
 &=-3g^2C_F\di^{\al_1}_{~\alpha_2}  \delta_{I}^{~J} \frac{1}{16 \pi^4 (\vec{x}^{\, 2}_{12})^2} \,.
\end{aligned}
\end{equation}
This result can be inserted in the diagram (a) in figure \ref{fig:gammascalars},  giving 
\begin{equation}\label{eq:ScalrPanela}
\begin{aligned}
\left.\langle \calD_{\rm SC}(\vec{x}_1)\calD_{\rm SC}(\vec{x}_2)\rangle^{(1)}\right|^{(a)}_{\rm log} & =\left.2\langle\phi_I^{\alpha_1}(\vec{x}_1) \phi^*{}^J_{\alpha_2}(\vec{x}_2)\rangle^{(0)}\langle\phi_J^{\alpha_2}(\vec{x}_1) \phi^*{}^I_{\alpha_1}(\vec{x}_2)\rangle^{(1)}\right|_{\rm{log }}\\&=- g^2 n_s \left(n_c^2-1\right) \frac{3}{16 \pi^6 (\vec{x}^{\, 2}_{12})^4} \,,
\end{aligned}
\end{equation}
where the factor 2 takes into account that both the upper and lower scalar propagators carry this contribution.

Let us now consider the diagram (b) in figure \ref{fig:gammascalars}. This evaluates to 
\begin{equation}\begin{aligned}
    \left.\langle \calD_{\rm SC}(\vec{x}_1)\calD_{\rm SC}(\vec{x}_2)\rangle^{(1)}\right|^{(b)}=-\frac{g^2}{2}  n_s (n_c^2-1) \intads dx\ dy\ & \Pi^{\mu\nu}(x,y) \left(K\left(x, \vec{x}_2\right) \overleftrightarrow{\nabla}^x_\mu K\left(\vec{x}_1, x\right)\right)\\&\left(K\left(y, \vec{x}_1\right) \overleftrightarrow{\nabla}^y_\nu K\left(y, \vec{x}_2\right)\right) \,,
  \end{aligned}  \label{eq:DscA}
\end{equation}
where $\overleftrightarrow{\nabla}_\mu=\overrightarrow{\nabla}_\mu-\overleftarrow{\nabla}_\mu$.
This expression can be rewritten as a linear combination of integrals in the form \eqref{eq:ref_int}. The result in $d=3$ reads
\begin{equation}
    \left.\langle \calD_{\rm SC}(\vec{x}_1)\calD_{\rm SC}(\vec{x}_2)\rangle^{(1)}\right|_{\rm log}^{(b)}\simeq g^2   n_s (n_c^2-1) \frac{5}{16 \pi^6 (\vec{x}^{\, 2}_{12})^4} \,.
\end{equation}
Let us now focus on the diagrams in which both $\calD_{\rm SC}$ and $\calD_{\rm YM}$ appear, i.e. the last two diagrams of figure  \ref{fig:gammascalars}. The diagram (c) is a contact diagram and therefore very simple to compute. It reads 
 \begin{equation}
    \left.\langle \calD_{\rm YM}(\vec{x}_1)\calD_{\rm SC}(\vec{x}_2)\rangle^{(1)}\right|^{(c)}=-\frac{g^2}{2}   n_s (n_c^2-1) \intads dx\ K^{i\nu}(\vec{x}_1,x) K_{i\nu}(\vec{x}_1,x) \ K\left(x, \vec{x}_2\right) ^2\,.
\end{equation}
Using again the procedure described in \cite{Ciccone:2024guw} for the computation of contact diagrams, we find
\begin{equation}
    \left.\langle \calD_{\rm YM}(\vec{x}_1)\calD_{\rm SC}(\vec{x}_2)\rangle^{(1)}\right|_{\rm log}^{(c)}\simeq -g^2  n_s (n_c^2-1) \frac{3}{4 \pi^6 (\vec{x}^{\, 2}_{12})^4} \,.
\end{equation}
The diagram (d), after Wick contractions, equals
\begin{equation}\label{eq:contFigdSc}
\begin{aligned}
    \langle \calD_{\rm YM}(\vec{x}_1)\calD_{\rm SC}(\vec{x}_2)\left.\rangle^{(1)}\right|^{(d)}=2g^2\,  n_s (n_c^2-1) & \intads dx\ dy\ K^{i\mu}(\vec{x}_1,x)K_i^{\ \nu}(\vec{x}_1,y)\\
    &\nabla^x_\mu K(x, \vec{x}_2) G(x,y)\nabla^y_\nu K(y,\vec{x}_2)\,,
    \end{aligned}
\end{equation}
where we integrated by parts and used that $\nabla_\mu^x K^{i\mu}(\vec{x}_1,x)=0$. 
The integral in \eqref{eq:contFigdSc} is of the kind of \eqref{eq:DscA} and can be evaluated in the same way. This gives
\begin{equation}
    \left.\langle \calD_{\rm YM}(\vec{x}_1)\calD_{\rm SC}(\vec{x}_2)\rangle^{(1)}\right|_{\rm log}^{(d)}\simeq g^2\,  n_s (n_c^2-1) \frac{5}{8 \pi^6 (\vec{x}^{\, 2}_{12})^4} \,.
\end{equation}
Putting all the results together, and rescaling $\calD_{\rm YM}$ as in the first of \eqref{eq:dispnorm} and $\calD_{\rm SC}$ as in \eqref{eq:dispnormS}, we get
\begin{equation}
\begin{split}
\Gamma_{22}& =-g^{-2}(\vec{x}^{\, 2}_{12})^4\left.\langle \calD_{2}(\vec{x}_1)\calD_{2}(\vec{x}_2)\rangle^{(1)}\right|_{\rm log}\simeq -\frac{(n_c^2-1)}{8n_c \pi^2} \,, \\
\Gamma_{12}&=\Gamma_{21}=-g^{-2}(\vec{x}^{\, 2}_{12})^4\left.\langle \calD_{1}(\vec{x}_1)\calD_{2}(\vec{x}_2)\rangle^{(1)}\right|_{\rm log}\simeq \frac{\sqrt{(n_c^2-1)n_s} }{8\sqrt{6n_c}\pi^2}\,,
\end{split}
\end{equation}
which reproduces \eqref{eq:matrixGsc12}, providing a non-trivial check of the result. 

\section{Physical implications}\label{sec:physical}

In this section we discuss some physical implications of the results found in sections \ref{sec:fermions} and \ref{sec:scalars}. We focus our analysis on the case of fermionic matter in the fundamental representation. This is at the same time the most interesting and the simplest case, since scalar matter unavoidably includes additional classically marginal deformations, the quartic self-interactions. Depending on the number of fundamental flavors (below the threshold of asymptotic freedom), the theory with fermionic matter flows in the IR to either a confining phase or a conformal phase. We will see that the anomalous dimensions of the lowest singlet scalar operators computed above provide useful information in both regimes. 

Regarding the confining phase, the discussion parallels the one for pure YM theory of \cite{Ciccone:2024guw}. As discussed in the introduction,
we expect that the Dirichlet bc should cease to exist beyond some critical value $L_\text{crit}$ of the AdS length also in the presence of fundamental matter.
We provide perturbative evidence that the merger and annihilation scenario advocated in \cite{Copetti:2023sya,Ciccone:2024guw} 
is a sensible explanation for the disappearance of the Dirichlet bc also in this case, and that the Neumann bc can smoothly reach the flat-space limit.

Regarding the conformal phase, we will discuss how the IR conformal symmetry implies the existence of a protected displacement operator at the fixed point, and we will show that, at least in the perturbative regime, this operator arises precisely from a linear combination of the scalar singlet operators that we studied above. This is a nontrivial check of our results.

\subsection{Confining phase}
The theory is expected to be in the confining phase for $0<n_f < n_f^*$, where the critical value $n_f^*<\frac{11}{2} n_c$ is an unknown function of $n_c$. In this phase, as remarked above, due to color confinement the Dirichlet boundary condition must disappear, while the Neumann boundary condition can approach the flat-space limit continuously. Barring some unexpected vanishing of OPE coefficients not explained by symmetry, merger and annihilation is guaranteed to happen whenever a boundary singlet scalar operator reaches dimension 3 \cite{Lauria:2023uca} (see also App.C of \cite{Ciccone:2024guw}). Starting from the perturbative regime at small $\Lambda L$, the leading candidates to reach this value of the scaling dimension are the lowest scalar singlets in the boundary spectrum. These are precisely the operators that start at dimension 4 in the free theory and whose anomalous dimensions were computed in section \ref{sec:fermions}.

The results in \eqref{eq:gammafinalF} show that the anomalous dimensions of the lowest singlets are both negative for the Dirichlet boundary condition, and both positive for the Neumann boundary condition. This is actually true not only for $0<n_f < n_f^*$ but also in the whole range of asymptotic freedom $0<n_f < \frac{11}{2}n_c$. Similar remarks apply to the anomalous dimensions \eqref{eq:gammafinalS} in the theory with scalar matter, for which asymptotic freedom holds in the range $0<n_s < 22 n_c$. In the confining phase, these results agree with the expectation that the Dirichlet boundary condition disappears by merger and annihilation, while the Neumann boundary condition extrapolates continuously to the flat-space limit. We will comment on their meaning in the conformal phase in the next subsection. The two negative anomalous dimensions for Dirichlet bc, and the two positive ones for Neumann bc, show that the theories with matter generalize the pattern found in the pure YM theory in \cite{Ciccone:2024guw} in the simplest possible way, and they also show the robustness of these indications coming from perturbation theory, making them less likely to be just a numerical coincidence.

With the Dirichlet bc, by looking at the most negative anomalous dimension, i.e. $\Delta_1^\text{D}$, we can estimate at which value of the AdS length $L_\text{crit}$ the transition occurs by a one-loop truncation. In terms of the flat-space dynamically generated scale $\Lambda$, we get
\begin{equation}\label{eq:estimate}
\Delta_{1}^{\rm D} = 3 ~~\Rightarrow~~(\Lambda L)_{\text{crit}}\vert_{\text{NLO}} = \exp\left({-\frac{ \gamma_1^\text{D}}{\beta_0}}\right)\,,
\end{equation}
where $\beta_0$ is defined in \eqref{eq:betagF}. Taking for instance $n_c=3$ and $n_f=0,1,2,3$, this gives $(\Lambda L)_{\text{crit}}\approx 0.37,0.33, 0.30, 0.27$ respectively. We see that the addition of matter makes the transition to occur for smaller values of $\Lambda L $, i.e. at smaller coupling, with respect to the pure gauge case.

\begin{figure}[t!]
    \centering
    \raisebox{-3.em}
 {\scalebox{1.1}{
    \begin{tikzpicture}
    \begin{scope}[shift={(4,6)}]
\draw (1,-5) [color=extragreen, rotate=90] parabola (2,0);
\end{scope}
\draw (0,2) [color=teal] parabola (5,3/2);
\draw (0,2) [color=cyan] parabola (7,3);
\draw (0,2) [color=blue] parabola (7,7/2);
    \begin{scope}[shift={(4,6)}]
\draw (1,-5) [dashed, color=red,rotate=90] parabola (0.27,-3);
\end{scope}
\draw [-stealth] (0,0) - - (0,3.5);
  \draw [-stealth] (-1,0) - - (7,0);
    \filldraw[fill=black, draw=black]  (5,1) circle (0.05 cm);
\draw (0,-1/2) - - (0,2);
\draw [dashed] (0,1) - - (5,1);
\draw [dashed] (5,0) - - (5,3.5);
 \node[below] at (7,0) {$\Lambda L $};
  \node[below] at (5.2,0) {$(\Lambda L )_\text{crit}$};
 \node[left] at (0,3.5) {$\Delta$};
  \node[left] at (0,2) {$4$};
   \node[left] at (0,1) {$3$};
\node[right] at (7.1,3.65)[color=blue]  {\scalebox{.9}{$\Delta_{1}^\text{N}$}};
\node[right] at (7.1,3.05)  [color=cyan]{\scalebox{.9}{$\Delta_{2}^\text{N}$}};
\node[right] at (5.1,1.6) [color=teal]  {\scalebox{.9}{$\Delta_{2}^\text{D}$}};
 \node[right] at (5.1,1)  [color=extragreen] {\scalebox{.9}{$\Delta_{1}^\text{D}$}};
\node[below] at (4.4,0.7)[color=red] {\scalebox{.9}{$\Delta_1^{\text{D}^*}$}};
\node at (4,4) {\scalebox{1}{Confining phase}};

\end{tikzpicture}}}
\caption{Schematic representation of the evolution of the scaling dimensions of the operators $\overline{\cal D}_{1,2}$ for both Dirichlet and Neumann bc, as a function of the AdS length $L$, in the confining phase. 
The red dashed line indicates the operator in the boundary conformal theory associated to the Dirichlet$^*$ bc, which becomes marginal at 
$L=L_\text{crit}$, where both Dirichlet and  Dirichlet$^*$ bc merge and disappear. The Dirichlet$^*$ bc must exist for $\Lambda L \lesssim (\Lambda L)_{\text{crit}}$, but it is not known how much it extends towards weak coupling.
\label{fig:DDstarN}}
\end{figure}
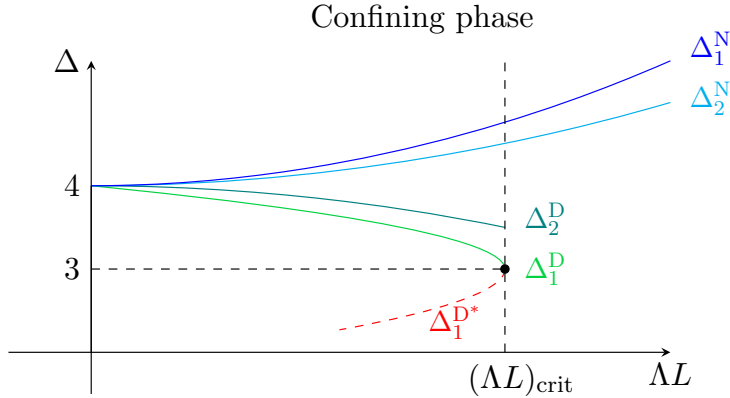

The picture that emerges in the confining phase is illustrated in figure  \ref{fig:DDstarN}. 

\subsection{Conformal phase} 

For $n_f^* \leq n_f < \frac{11}{2}n_c$ the theory is expected to be in the conformal phase, i.e. to flow to an interacting conformal field theory in the IR. As we increase the AdS radius $L$, the coupling $g^2$ at the scale $\frac{1}{L}$ starts increasing from the initial value $g^2\ll 1$ at small $L$. For large $L$ it settles at the fixed point value $g^2_*$. At the fixed point $g^2_*$, thanks to conformal symmetry, the theory in AdS can be mapped via a Weyl rescaling to a flat-space BCFT, i.e. it is equivalent to the CFT in half flat space $\mathbb{R}^3 \times\mathbb{R}_+$ with a conformal boundary condition. This means that, by studying the conformal phase in AdS, we can also learn about properties of its flat-space conformal boundary conditions. 

As we already remarked above, a universal property of BCFTs is the existence of a boundary scalar operator of protected scaling dimension equal to the bulk spacetime dimension, i.e. $\Delta=4$ in our setup, the displacement operator \cite{McAvity:1993ue, McAvity:1995zd}. In section \ref{sec:displop} we focused on the displacement operators of the free UV BCFT, there were two of them because of the decoupling between the matter and YM sector. Those operators stopped being protected as a consequence of the interaction and their scaling dimension moved away from $4$. The observation we make here is that in the conformal phase an operator of dimension $\Delta=4$ should again appear in the boundary spectrum in the IR, as a consequence of the conformal symmetry. In this case only one is expected, because there is only one interacting sector. 

It might seem difficult to test this prediction using our calculations because we used perturbation theory, and the CFT in the IR is typically strongly coupled. However, even the IR conformal phase admits a perturbative treatment if we consider the Veneziano limit
\be
n_c\to \infty\,,~~n_f\to \infty\,,~~g^2\to 0\,,~~~\text{with}~~~\lambda \equiv \frac{g^2 n_c}{16 \pi^2}~~\text{and}~~x\equiv\frac{n_f}{n_c}~~~\text{fixed}~.
\ee
The $\beta$-function of the theory up to two loops in this limit reads
\be
\beta^\text{Ven} = \frac{4\lambda^2}{3}\Big(x-\frac{11}{2}\Big) + \frac{2\lambda^3}{3}(13x-34)+{\cal O}(\lambda^4)\,.
\label{eq:betaVen}
\ee
One can then use the continuous parameter $x$ to establish rigorously the existence of the conformal phase, and to set up a systematic perturbative expansion, by further taking the Banks-Zaks limit \cite{Banks:1981nn}, i.e. $x=11/2-\epsilon$, with $\epsilon\ll 1$. The $\beta$-function \eqref{eq:betaVen} has a fixed point at 
\be\label{eq:lambdaBZ}
\lambda_\text{BZ} = \frac{4}{75} \epsilon + {\cal O}(\epsilon^2)\,,
\ee
which is parametrically under control for small $\epsilon$.

An interesting property of the one-loop anomalous dimension \eqref{eq:gammafinalF} is that the one with smallest absolute value, i.e. $\gamma_2$, vanishes at the upper edge of the conformal window $n_f = \frac{11}{2}n_c$. Note that the same is true for the theory with scalar matter, for which the upper edge is $n_s = 22 n_c$, see  \eqref{eq:gammafinalS}.\footnote{The fact that the anomalous dimension vanishes when we are at a fixed point with $\beta =0$ naively looks like an immediate consequence of \eqref{eq:AllorderRel}. However, it is still possible to satisfy \eqref{eq:AllorderRel} with $\Delta_i \neq d+1$, as long as the coefficient $C_{T\overline{\mathcal{D}}_i}$ vanishes at the fixed point. Actually, if we have an IR interacting fixed point, generically we expect only one among the possibly many operators $\overline{\mathcal{D}}_i$ to provide the displacement in the IR. All the others are instead expected to satisfy $C_{T\overline{\mathcal{D}}_i} = 0$ at the fixed point. Indeed we will see below that in the case of QCD the operator $\overline{\mathcal{D}}_1$ is of this second type, having $\Delta_1 \neq 4$.} The Banks-Zaks expansion parameter $\epsilon$ is precisely the distance of $x$ from the upper edge of the conformal window. This implies that, in the Veneziano limit, the leading order $\mathcal{O}(\lambda)$ anomalous dimension of $\overline{\cal D}_{2}$ has a coefficient that scales like $\mathcal{O}(\epsilon)$ in the limit $\epsilon \to  0$. Therefore, when we plug the fixed-point value of the coupling \eqref{eq:lambdaBZ}, we find a result for the anomalous dimension that is negligible at this order, namely
\be\label{eq:dispIR}
\Delta_{2}^{\text{D/N}} = 4 + {\cal O}(\epsilon^2)\,.
\ee
At the order we are computing, namely $\mathcal{O}(\epsilon)$, this is consistent with the BCFT prediction of the existence of a protected displacement operator. From the general argument, we expect that the scaling dimension will remain vanishing at any order in $\epsilon$, but one needs calculations at higher loops to check the order ${\cal O}(\epsilon^2)$ and above. Note that \eqref{eq:dispIR} holds both for Neumann and Dirichlet bc, because we derived it from the vanishing of $\gamma_2 = \gamma_2^\text{D} = -\gamma_2^\text{N}$ for $n_f = \frac{11}{2}n_c$.

For $\overline{\cal D}_1$ we have
\be\label{eq:D1eps}
\Delta_{1}^{\text{D/N}} = 4 \mp \frac {8}{15} \epsilon+ {\cal O}(\epsilon^2)\,,
\ee
where the $-$ and the $+$ correspond to Dirichlet and Neumann bc, respectively. 
Via Weyl rescaling, this result provides the scaling dimensions of the lightest (second lightest) singlet operator, for the Dirichlet (Neumann) BCFT of the conformal window of QCD, in the Banks-Zaks expansion.

For Dirichlet boundary conditions, the operator $\overline{\cal D}_1$ is the one with the largest negative anomalous dimension that we proposed to cause merger and annihilation in the confining phase. Of course its anomalous dimension in the IR conformal phase is small as long as $\epsilon \ll 1$ and the perturbative expansion is reliable. In the next subsection we will speculate about extrapolating it beyond small values of $\epsilon$, and about the possible significance of $\Delta_{1}^{\text{D}}$ crossing marginality as a function of $\epsilon$.

\begin{figure}[t!]
    \centering
    \raisebox{-3.em}
 {\scalebox{1.1}{
    \begin{tikzpicture}[domain=0:7]
\draw  [color=cyan] (0,2) - - (7,2);
\draw  [color=teal,dashed] (0,2) - - (7,2);
\draw  [color=teal,dashed] (0,2) - - (7,2);
\draw  [dashed] (0,3/4) - - (7,3/4);
\draw [-stealth] (0,-1) - - (0,3);
  \draw [-stealth] (-1,0) - - (7,0);

\draw (0,-1) - - (0,2);
  \draw[color=blue,samples=100]    plot (\x,{43/22+0.5/(1+100^(0.5-0.5*\x))})  node[right]{$\,$};        
  \draw[color=extragreen,samples=100]    plot (\x,{45/22-0.5/(1+100^(0.5-0.5*\x))})  node[right]{$\,$}; 
 \node[below] at (7,0) {$\Lambda L $};
 \node[left] at (0,3) {$\Delta$};
  \node[left] at (0,2) {$4$};
\node[left] at (0,3/4) {$3$};
\node[right] at (5.4,2.8)[color=blue]  {\scalebox{.9}{$\Delta_{1}^\text{N}$}};
\node[right] at (7.1,2)  {\scalebox{.9}{$\Delta_{2}^\text{N,D}$}};
 \node[right] at (5.4,1.2)  [color=extragreen] {\scalebox{.9}{$\Delta_{1}^\text{D}$}};
\node at (4.1,3.5) {\scalebox{1}{Conformal phase}};
\end{tikzpicture}}}
\caption{Schematic representation of the evolution of the scaling dimensions of the operators $\overline{\cal D}_{1,2}$ for both Dirichlet and Neumann bc, as a function of the AdS length $L$, in the conformal phase with $x>x_0$. 
The horizontal black line represents the displacement operator present for both Dirichlet and Neumann bc.
With a short RG flow, the scaling dimension of the operator $\overline{\cal D}_{1}$ approaches the IR fixed point value as $L$ is increased.
Possible scenarios for $x_* \leq x \leq x_0$ are discussed in subsection \ref{subsec:endcw}.
\label{fig:DDstarNBZ}}
\end{figure}
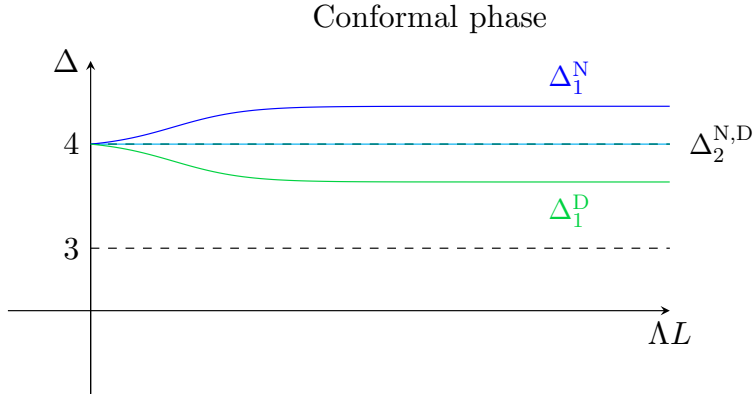

The picture emerging in the conformal phase is illustrated in figure  \ref{fig:DDstarNBZ}.

\subsection{End of the conformal window}\label{subsec:endcw}
Merger and annihilation is a compelling scenario for the end of the (bulk) conformal window in QCD theories \cite{Kaplan:2009kr} 
(see also \cite{Gies:2005as}).
This scenario requires that, at least close to the lower edge $x_* = \frac{n_f^*}{n_c}$ of the conformal window, there exists another fixed point, dubbed QCD$^*$. Conformality is lost because the fixed points associated to QCD and QCD$^*$, which have the same global symmetries, approach each other until they merge and annihilate. Notably, at the merging point, a bulk operator, namely the one governing the RG flow from one fixed point to the other, becomes marginal.

In this section we speculate on how the bulk merger and annihilation responsible for the end of the conformal window could be detected using the boundary CFT data of the theory in AdS, and in particular about its possible relation to the disappearance of the Dirichlet bc. 

To this end, let us consider the scaling dimension $\Delta_1^{\text{D}}(\lambda , x)$ of the lowest scalar singlet
with Dirichlet bc, as a function of the 't Hooft coupling $\lambda$ at the AdS scale $1/L$, and of the parameter $x$. 
For any fixed $x$ in the conformal window, we can ask about the behavior of the function $\Delta_1^{\text{D}}$ as we increase the radius. By continuity and consistency with the perturbative result \eqref{eq:D1eps}, when $x<11/2$ is sufficiently close to $11/2$ this dimension will reach a value $\Delta_1^{\text{D}}(\lambda_\text{BZ} , x)> 3$. This gives the boundary scaling dimension for $\overline{\mathcal{D}}_1$ in the corresponding BCFT of the IR fixed point.
The most natural expectation for the behavior of $\Delta_1^{\text{D}}(\lambda_\text{BZ} , x)$ as a function of $x$ is that it starts from the displacement value $4$ at $x=11/2$ and decreases monotonically as $x$ is decreased. There are then two possibilities:
\begin{itemize}
\item[(i)]{For some value $x_0$, with $x_* < x_0 < 11/2$, the dimension reaches the marginal value, i.e. $\Delta_1^{\text{D}}(\lambda_\text{BZ} , x_0) = 3$. If we keep decreasing $x$ to values in the range $x_* < x < x_0$,  the Dirichlet boundary condition is expected to disappear as a real conformal boundary condition of the bulk CFT. It still exists as a complex BCFT, in which $\Delta_1^{\text{D}}(\lambda_\text{BZ} , x)$ is a complex number;}
\item[(ii)]{At the edge of the conformal window we still have a real value $\Delta_1^{\text{D}}(\lambda_\text{BZ} , x_*) \geq 3$.}
\end{itemize}

Note that, as soon as $x<x_*$, the theory is in the confining phase, therefore the scaling dimension of the lowest singlet is expected to cross marginality as we increase the radius, causing the Dirichlet bc to disappear as a real, AdS invariant bc. 
If we insist on following the Dirichlet bc, then, as $x$ crosses $x_*$ we expect $\Delta_1^{\text{D}}$ to reach a complex value at large radius. 

As a result, if one assumes continuity in the large-radius value of this scaling dimension across the bulk transition at $x=x_*$, one is led to the conclusion that the option (i) is realized.\footnote{If one further assumes that the disappearance of the Dirichlet bc is in one-to-one correspondence with confinement, then the point $x_0=x_*$ is uniquely singled out.} This opens the possibility of constraining $x_*$ using the data of the boundary CFT: as we discussed in the previous subsection, the function $\Delta_1^{\text{D}}(\lambda_\text{BZ} , x)$ can be computed in a Banks-Zaks perturbative expansion in $\epsilon=11/2-x$, and it can be then used to estimate the value $x_0$ and get an inequality $x_* \leq x_0$. 

We do not have strong arguments in favor of this continuity assumption. It is nevertheless interesting to notice that with the logic outlined above, using a drastic one-loop truncation of \eqref{eq:D1eps} leads to the estimate 
\be\label{eq:xstar}
x_* \leq  \frac{29}{8}\approx 3.6\,,
\ee
which compares well with other studies \cite{Appelquist:1996dq,Jarvinen:2011qe,DiPietro:2020jne}. We can also use the same logic outside of the strict Veneziano limit, simply by working at finite $n_c$ and treating $n_f$ as a continuous parameter. Doing so, the analogous estimate for ordinary QCD with $n_c=3$ gives 
\be\label{eq:nfstar}
n_f^* \leq \frac{5757}{524}\approx 11\,.
\ee
This estimate again compares favorably with known results.\footnote{For QCD, analytical and lattice studies provide evidence that $n_f^*< 12$, see \cite{Hasenfratz:2024fad} and references therein, and other lattice studies indicate that $n_f^*> 6$ \cite{LSD:2009yru,Miura:2011mc}.}

Since we are proposing a novel approach to estimate the end of the conformal window based on the conformal data at the boundary, it seems appropriate to compare it with more standard approaches based on the bulk data of the CFT. As we will see, with the available data these attempts fail at giving reasonable estimates, which is likely the reason why we could not find them in the literature. As briefly reviewed above, similarly to the disappearance of the Dirichlet bc, also the disappearance of the bulk conformal window is expected to be triggered by merger and annihilation. Hence it should also require a scalar singlet operator, this time a bulk one, to reach marginality, i.e. scaling dimension $4$. One can then look at the perturbative calculations, within the Banks-Zaks expansion, of the dimensions of operators at the IR fixed point, and use them to estimate when these operators reach marginality. The leading candidates are operators of dimension $6$ in the free UV theory. Moreover, in the Veneziano limit, large $N$ factorization requires that the operator becoming marginal at the fixed point is a double trace operator \cite{Gubser:2002vv}. These requirements single out four-fermion flavor singlet operators. There are four such operators, $O_1 = (\bar\psi \gamma^\mu \psi)^2$,
$O_2 = (\bar\psi \gamma^\mu \gamma_5 \psi)^2$, $O_3 = (\bar\psi \gamma^\mu t^a \psi)^2$, $O_4 = (\bar\psi \gamma^\mu \gamma_5 t^a \psi)^2$.
These operators mix under renormalization and their anomalous dimension matrix is known at one-loop order, see e.g. \cite{Beneke:1997qd,Bauer:1997gs}. The only operator $O$ with a negative eigenvalue $\gamma$ at one loop is a combination of $O_3$ and $O_4$, with $\gamma = -\epsilon (13+\sqrt{565})/3$. If we truncate the scaling dimension keeping only this leading order $\mathcal{O}(\epsilon)$, we conclude that this operator remains irrelevant for any $x\geq 0$, failing to reach marginality (marginality is only reached at the negative value $x_*\approx -4.95$). A similar estimate can be done for standard QCD, by keeping $n_c=3$ and treating $n_f$ as a continuous parameter. In this case all four operators mix, but again, if we truncate at one loop, even the lowest dimensional of them fails to reach marginality at any $n_f \geq 0$ (marginality is reached at the negative value $n_f^* \approx -10$). 
These negative results show that a prediction of the lower edge of the conformal window is hardly attainable with perturbative methods at low orders (see \cite{DiPietro:2020jne} for a more extensive discussion). This is a reason to be skeptical of the one-loop estimates \eqref{eq:xstar}  and \eqref{eq:nfstar}, and suggests that they could be mere numerical coincidences. A more optimistic possibility is that the boundary data has better convergence properties.

\section{Chiral symmetry breaking}\label{sec:chisb}

So far in this paper we have been focusing on the extension of the results of \cite{Ciccone:2024guw} about confinement in pure YM theory in AdS to the theory with matter, and in addition we have studied the conformal phases from the point of view of AdS. Notably, besides the existence of conformal windows, there is another new phenomenon that can occur in gauge theories due to the presence of matter: the spontaneous breaking of global chiral symmetries. 
Remarkably, chiral symmetry breaking manifests already at tree-level in AdS because of the necessity of imposing boundary conditions on the fermions. In this section we develop a bit more on this topic. More specifically, in subsection \ref{subsec:fermbc} we discuss more general
fermion boundary conditions, for both massless and massive fermions. 
In subsection \ref{subsec:FSlimit} we develop a bit on the flat-space limit and speculate on how the perturbative AdS chiral symmetry breaking can be connected to the non-perturbative flat-space limit, taking also into account the constraints found in \cite{Vafa:1983tf}.

\subsection{General fermion boundary conditions and symmetries}
\label{subsec:fermbc}
As explained in section \ref{sec:fermions}, assigning boundary conditions to fermions in AdS leaves the freedom of choosing the matrix $B$ in \eqref{eq:fermbc}. This matrix must commute with the mass matrix $M$ (if nonzero) and have eigenvalues
$\pm1$.  So far, we have adopted the maximally symmetric choice $B_I^{~J}=\pm \delta_I^{~J}$. We now turn to more general boundary conditions and provide the motivation for our choice.

To this end, let us discuss the global symmetries and their interplay with the boundary conditions.
Setting $M=0$ and $d=3$, the theory of $n_f$ fermions in the fundamental representation of $SU(n_c)$ gauge group has the following global 0-form symmetry in the bulk
\begin{equation}\label{eq:bulkSymm}
    G=\frac{SU(n_f)_L\times SU(n_f)_R\times U(1)_V}{\ZZ_{n_c}\times \ZZ_{n_f}}~.
\end{equation}
Here $SU(n_f)_{L,R}$ are the chiral flavor symmetries acting on the Weyl components of the Dirac fermion, and $U(1)_V$ is the vector-like baryon number symmetry. Denoting an element of $SU(n_f)_L\times SU(n_f)_R\times U(1)_V$ as $(V_L, V_R, e^{i \alpha})$, the discrete subgroups that we are quotienting by are those generated by the elements
\begin{align}
\begin{split}
&\mathbb{Z}_{n_c}~:~~(\mathds{1}, \,\mathds{1},\, e^{-i \frac{2\pi}{n_c}})~,\\
& \mathbb{Z}_{n_f}~:~~(e^{i \frac{2\pi}{n_f}}\mathds{1},\, e^{i \frac{2\pi}{n_f}}\mathds{1},\, e^{-i \frac{2\pi}{n_f}})~.
\end{split}
\end{align}
The quotient is needed to eliminate subgroups that act trivially on any gauge-invariant operator. Adding a mass term in the bulk breaks explicitly part of this global symmetry, for instance $M_I^{~J} = m \,\delta_I^{~J}$ with $m\neq 0 $ breaks the symmetry to $U(n_f)_V/\mathbb{Z}_{n_c}$. So far this discussion has been about the bulk symmetry and it is not specific to the AdS background. The additional interesting question to ask in AdS is what symmetry acts on the boundary conformal correlators. The matrix $B_I^{~J}$ appearing in the boundary condition can further reduce the symmetry group at the boundary. Given that this breaking only takes place at infinity, it should be understood as a {\it spontaneous} breaking of the symmetry. The constraint we found from AdS isometries, namely that $B_I^{~J}$ must commute with $M_I^{~J}$, is consistent with this interpretation: only the subgroup of the global symmetry $G$ that is not explicitly broken by the mass is an actual symmetry of the theory, that can undergo spontaneous breaking. Note that, unlike the mass, the matrix $B_I^{~J}$ cannot be set to zero because its eigenvalues are $\pm 1$. Therefore, for $M=0$, even at arbitrarily small coupling the symmetry $G$ is broken spontaneously in AdS. The most symmetric possible choice for $B_I^{~J}$ in this case is $\pm \delta_I^{~J}$, and its orbit under the action of the broken generators. With this choice, the symmetry acting on the boundary correlators in the massless theory is 
\begin{equation}\label{eq:boundarySymm}
    \widehat{G}^{\text{D}} =\frac{U(n_f)_V\times SU(n_c)}{\ZZ_{n_c}} \,, \qquad\qquad
       \widehat{G}^{\text{N}} =\frac{U(n_f)_V}{\ZZ_{n_c}} \,.
\end{equation}
where D and N stand, respectively, for the choice of Dirichlet or Neumann bc for the gauge fields.\footnote{
The form of the global symmetries \eqref{eq:bulkSymm} and \eqref{eq:boundarySymm} depend on the choice of representation for the matter field. More in general, 
for a representation with $n$-ality $r$ under $SU(n_c)$, (i) an electric one-form symmetry $\ZZ_{\text{gcd}(r,n_c)}^{(1)}$ remains unbroken in $G$ and $\widehat{G}^{\text{N}}$; (ii) the $\ZZ_{n_c}$ in the quotient should be replaced by $\ZZ_{n_c/\text{gcd}(r,n_c)}$ in $G$, $\widehat{G}^{\text{D}}$ and $\widehat{G}^{\text{N}}$.} These groups are given by the bulk symmetry that is not spontaneously broken by the boundary conditions, with possibly in addition an  asymptotic symmetry coming from the bulk gauge fields, if they satisfy Dirichlet bc. This second type of symmetry, unlike the first, acts locally on the boundary, namely there are boundary conserved currents associated to it.

In analogy with the Goldstone theorem in flat space, the consequence of the spontaneous breaking of continuous symmetries in AdS$_{d+1}$ is the existence of scalar operators of protected dimension $d$, the so-called {\it tilt} operators $t(\vec{x})$ (see \cite{Bray:1977fvl, Herzog:2017xha, Cuomo:2021cnb, Padayasi:2021sik} for discussions in the BCFT context, adapted to QFT in AdS in \cite{Copetti:2023sya}). These operators appear in the boundary OPE of the conserved currents associated to the broken symmetries, i.e. $J_z(z,\vec{x})\underset{z\to 0}{\sim} z^{d-1} \,t(\vec{x})$. We can easily verify this statement in the theory we are considering, at least in the limit of weak coupling $g^2\to 0$. For instance, if we consider a $2$-dimensional flavor subspace in which the mass matrix is $m \mathds{1}_{2\times 2}$, giving a $U(2)$ flavor symmetry, we can have either a boundary condition with $B = \text{diag}(\pm 1 , \pm 1)$ in this block, or with $B = \text{diag}(\pm 1 , \mp 1)$. In the first case, $B$ does not break any symmetry that is not already explicitly broken by the mass, the boundary operators $\widehat{\Psi}_{\pm\,I=1,2}$ have the same scaling dimension $\Delta_{1,2} = \frac{d}{2}\pm m$ and there is no composite scalar operator of dimension $d$. Viceversa, in the second case the boundary condition spontaneously breaks $U(2)$ to $U(1)\times U(1)$, the boundary operators are $\widehat{\Psi}_{\pm\,1}$ and $\widehat{\Psi}_{\mp\,2}$ with respective dimensions $\Delta_1 = \frac{d}{2}\pm m$ and $\Delta_2 = \frac{d}{2}\mp m$, and the two off-diagonal currents $\bar{\Psi}^1 \gamma_\mu \Psi_2$ and $\bar{\Psi}^2 \gamma_\mu \Psi_1$ are such that their $z$ components at the boundary give the composite operators $\widehat{\bar{\Psi}}{}_{\pm}^1\widehat{\Psi}_{\mp\,2}$ and $\widehat{\bar{\Psi}}{}_{\mp}^2\widehat{\Psi}_{\pm\,1}$ of dimension $d$, i.e. the tilt operators. While this is only a simple check in the free limit, the prediction from the spontaneously broken symmetry is that these operators remain of dimension $d$ at finite $g^2$. In the massless case, all the flavor components have the same scaling dimension $\frac{d}{2}$, therefore all the gauge-invariant non-vanishing bilinears at the boundary potentially give rise to tilt operators. Specifying again to AdS$_4$, and taking the maximally symmetric boundary condition with symmetry group \eqref{eq:boundarySymm}, we see that the unbroken currents $(J^V_\mu)^I_{~J} = \bar{\Psi}^I\gamma_\mu \Psi_J$ are such that their $z$ component vanishes at the boundary, as a consequence of the boundary condition, while the broken ones $(J^A_\mu)^I_{~J} = (\bar{\Psi}^I \gamma_5\gamma_\mu \Psi_J - \text{trace})$ give rise to tilt operators. Note that the current $J^A_\mu = \bar{\Psi}^I \gamma_5\gamma_\mu \Psi_I$ of the axial $U(1)$ is broken by the
Adler-Bell-Jackiw anomaly \cite{Adler:1969gk,Bell:1969ts}, and as a result we expect that its $z$ component at the boundary, while it is a non-vanishing operator which has dimension $d$ in the free limit, will not be protected and will get an anomalous dimension as we turn on the interaction $g^2$.\footnote{See \cite{Copetti:2025sym} for a recent discussion, in the context of BCFT, of the effect of bulk anomalies on the family of boundary conditions parametrized by tilt operators.}

Summarizing the discussion of symmetries, we have seen that at weak coupling there is an additional parameter $B^I_{~J}$ determining the bc for the fermions, and depending on this parameter we can have different patterns of spontaneous symmetry breaking of the bulk symmetry, resulting in different residual symmetries acting on the boundary. Any choice of $B^I_{~J}$ belongs to a continuous family of bc with the same symmetry, related to each other by the action of the broken generators. From the point of view of the boundary conformal theory, we can move in this continuous family by deforming the theory with the exactly marginal tilt operators, which have the effect of modifying the boundary term and the associated boundary condition. In the rest of the paper, we focused on the massless case and the maximally symmetric bc preserving \eqref{eq:boundarySymm}. These bc have the same pattern of spontaneous symmetry breaking as the one we believe to be realized in flat-space QCD, with the difference that in AdS it is also visible at weak coupling.

\subsection{Flat-space limit}
\label{subsec:FSlimit}

In flat space, in gauge theories coupled to matter, such as QCD, both confinement and chiral symmetry breaking are non-perturbative phenomena,
possibly related in a way we still do not fully understand. 
In contrast, the two phenomena appear quite differently when QCD is defined on AdS. Confinement due to the AdS curvature occurs at weak coupling when Neumann bc are imposed. With Dirichlet bc, instead, we conjecture that it is responsible for the disappearance of the bc itself, like in the pure YM case  \cite{Aharony:2012jf,Ciccone:2024guw}.
In contrast, chiral symmetry breaking appears at tree-level independently of the choice of bc for both gauge and fermion fields, as discussed in section \ref{subsec:fermbc}. Such breaking can be quantitatively computed by taking the coincident-point limit of the propagator \eqref{eq:propfbulk}. After subtracting a normal-ordering UV divergence, we find the following 1-point function of the gauge-invariant fermion bilinear
\begin{equation}\label{eq:TLChiSB}
     \langle \bar\Psi^J_\al \Psi^\al_I\rangle =-B_I^{~J}\frac{n_c}{4\pi^{2} L^{3}}\,,
\end{equation}
where we have reinstated the $L$ dependence to stress that it vanishes in the flat-space limit. The fact that this condensate is non-zero and is aligned with the parameter $B_I^{~J}$ of the boundary condition agrees with the interpretation of $B_I^{~J}$ as spontaneous symmetry breaking. We see that this condensate is present in AdS even for $g^2 = 0$. 

This striking difference should not come as a surprise. As is well-known, QCD has a plethora of 't Hooft anomalies involving the flavor group \eqref{eq:bulkSymm}
\cite{Frishman:1980dq}. It is also well-known that a QFT defined on a manifold with boundary cannot admit $G$-preserving boundary conditions if the group $G$ has 't Hooft anomalies, see e.g. \cite{Han:2017hdv, Thorngren:2020yht}. This is at the origin of the impossibility of having fermion boundary conditions preserving the whole group \eqref{eq:bulkSymm}, as extensively discussed in the previous section, and hence at the origin of the tree-level chiral symmetry breaking \eqref{eq:TLChiSB}.
Higher-order perturbative corrections would lead to the following asymptotic series for the quark condensate
for $d+1=4$:
\begin{equation}\label{eq:AllOrderChiSB}
     \langle \bar\Psi^J_\al \Psi^\al_I\rangle \sim B_I^{~J}\frac{1}{L^{3}} \sum_{n=0}^\infty c_n g^{2n}\,,
\end{equation}
where $g^2=g^2(1/L)$, $c_0$ can be read from \eqref{eq:TLChiSB} and $c_{n>0}$ are higher-order coefficients coming from radiative corrections.
When the condensate is computed with Neumann bc for the gauge field, the bc is expected to persist continuously up to $L=\infty$. It is then reasonable to expect that \eqref{eq:AllOrderChiSB} applies for any $L$ up to $L=\infty$.
In what follows we then discuss how the flat-space limit of the perturbative chiral symmetry breakings \eqref{eq:AllOrderChiSB} relates to the known results in flat space.

We start by considering the pure axial breaking, with 
\be\label{eq:axialBreak}
B_I^{~J}=\pm\delta_I^{~J}\,.
\ee
At first glance, it is not obvious how \eqref{eq:AllOrderChiSB}, with $B_I^{~J}=\pm\delta_I^{~J}$ and in the $L\to \infty$ limit, can lead to the known non-perturbative condensate occurring in flat space:
\begin{equation}\label{eq:NPChiSB}
     \langle \bar\Psi^J_\al \Psi^\al_I\rangle = \delta_I^{~J} c \Lambda^3 \,,
\end{equation}
where $c$ is a constant and $\Lambda$ is the dynamically generated scale $\Lambda\approx \mu \exp(-1/(\beta_0 g^2(\mu))$.
When $L\to \infty$, indeed, \eqref{eq:AllOrderChiSB} vanishes, suggesting naively that the non-perturbative flat-space breaking \eqref{eq:NPChiSB} is unrelated to the AdS perturbative one. We would like to argue that, instead, \eqref{eq:AllOrderChiSB} can be related to \eqref{eq:NPChiSB}
in an intriguing way. 

In order to explain the key point, it is useful to consider a quantum mechanical example with similar features, where the connection has been 
firmly established. This is the supersymmetric double well, one of the simplest models featuring dynamical supersymmetry breaking \cite{Witten:1981nf}.
The bosonic potential of the supersymmetric double well equals the one of the ordinary symmetric double well. Once fermions are integrated out, we get an additional term, linear in the bosonic variable \cite{Balitsky:1985in}. The coefficient of this term is such that the ground state energy vanishes to all orders in perturbation theory. Non-perturbatively, the ground state has a positive energy, detectable for instance from an instanton computation.
Despite the striking property of having a totally vanishing perturbative contribution, it turns out that we can reproduce the non-perturbative ground state energy using perturbation theory. In fact, two ways have been discussed in the literature. Both methods feature deformations which explicitly break supersymmetry and hence give rise to a non-trivial perturbative asymptotic series for the ground state energy. 

In the first class of deformations \cite{Serone:2016qog,Serone:2017nmd}, the deformed potential is such that the perturbative asymptotic series is ensured to be Borel resummable. The key point is that if we first Borel resum the perturbative result and after we remove the deformation, we end up reproducing the exact ground state energy. Such expansions have been dubbed exact perturbation theory (EPT) in \cite{Serone:2016qog,Serone:2017nmd}.

In the second class of deformations \cite{Kozcaz:2016wvy}, the deformed potential gives rise to a non-Borel resummable perturbative asymptotic series.
One can then study the Stokes phenomena of the deformed series, using resurgence techniques, to predict the missing deformed non-perturbative terms.
The key point here is that if we remove the deformation after the resurgence analysis, the perturbative series vanishes, as it should, but the
non-perturbative one does not, and reproduce the non-perturbative contribution one would get from instantons. 
Such mechanism has been dubbed Cheshire cat resurgence in \cite{Kozcaz:2016wvy}.

The analogy of this example with chiral symmetry in AdS should now be clear. The undeformed theory corresponds to flat space, $L=\infty$, in which case
the quark condensate is only non-perturbatively generated. Putting the theory to an AdS with finite $L$ corresponds to a deformation which generates 
the asymptotic series \eqref{eq:AllOrderChiSB}. In contrast to the quantum mechanical case where we could choose which class of deformations to make, 
AdS corresponds to a specific deformation and it is not known if this is of the first or second class discussed above.
Depending on this, the series \eqref{eq:AllOrderChiSB} will turn out to be Borel or non-Borel resummable.\footnote{In flat space, asymptotic series in gauge theories are generally non-Borel resummable due to the presence of IR renormalon singularities. It is well possible that these are tamed in AdS by the presence of the IR cut-off $1/L$, with other possible singularities of different origin being effectively negligible.} In both cases, the knowledge of the $c_n$ would allow us to reproduce the non-perturbative flat-space condensate using directly the series (first class) or its resurgent properties (second class).

Having speculated on how flat-space chiral symmetry breaking can be related to the AdS perturbative one, we now consider the non-pure axial breakings with 
\be
B_I^{~J}\neq \pm\delta_I^{~J}\,.
\ee
In this case, the vector $SU(n_f)_V$ global symmetry gets generally broken to a subgroup, and
we have to face another possible puzzle. In flat space, it is known that in non-abelian gauge theories with a non-chiral fermion spectrum (vector-like theories), no Yukawa couplings, and at $\theta=0$, vector global symmetries cannot be broken \cite{Vafa:1983tf}.
The findings of \cite{Vafa:1983tf} do not directly apply for spaces with boundaries such as AdS, and hence do not forbid the presence of the condensates \eqref{eq:AllOrderChiSB} with $B_I^{~J}\neq \pm\delta_I^{~J}$ for generic values of $L$. However, 
they imply that out of all the possible bc for the fermions at weak coupling, only the ones in \eqref{eq:axialBreak} can exist when AdS is large and we approach the flat-space limit. 
The bc with $B_I^{~J}\neq \pm\delta_I^{~J}$, like the Dirichlet bc for gauge fields, 
should then stop existing beyond some critical length $L$ due to some mechanism, such as merger and annihilation.

\section{Outlook}\label{sec:outlook}

Several open questions and promising avenues for future research emerge from this paper.
We have found evidence for the disappearance of the Dirichlet bc in the confining phase, while in the Banks-Zaks regime we have observed the existence of a displacement operator at the boundary, in agreement with conformality in the bulk. Under the assumption of continuity of the boundary data at large radius, we have proposed a novel way to bound the lower edge of the conformal window. Remarkably, the perturbative estimates obtained in this way, even at a one-loop level truncation, show good agreement with results available in the existing literature. Given the interest and efforts in the determination of $n_f^*$, it would be interesting to compute the anomalous dimension of $\Delta_1^{\text{D}}$ at two-loop order. In this way we can improve the bound \eqref{eq:nfstar} and verify its stability.

Our new proposal to determine endpoints of conformal windows can be used in setups different from QCD$_4$, such as 3d gauge theories with an even number $n_f$ of massless fermions. It is known that when $n_f$ is sufficiently large, both in the abelian and non-abelian case, the RG flow leads to an interacting CFT in the IR \cite{Pisarski:1984dj,Appelquist:1988sr,Appelquist:1989tc}. However, both the existence and the value of the endpoint of the conformal window have been
longstanding points of controversy. For QED$_3$ there is good evidence that $n_f=4$ is conformal while  $n_f=2$ is not (see e.g. \cite{Rychkov:2023wsd} for a review), but for QCD$_3$ the problem is still open (see \cite{DeCesare:2022obt} for some recent estimate). 
A similar question exists for $O(N)\times O(2)$-symmetric models in 3 dimensions \cite{HikaruKawamura_1998,Delamotte:2003dw,Delamotte:2010ba}, where the conformal bootstrap suggests $N^*=4$ \cite{Reehorst:2024vyq}. 

Chiral symmetry breaking in AdS is another interesting line for future investigations. We have argued that
a class of fermion boundary conditions should stop existing at large $L$ in order not to conflict with the Vafa-Witten theorem \cite{Vafa:1983tf}.
The disappearance at large $L$ of boundary conditions (through merger and annihilation) with spontaneous symmetry breaking patterns that are not realized in flat space was already seen in \cite{Copetti:2023sya} in the context of 2d models and at large $N$. 
It would be interesting to establish this disappearance in 4d QCD and to understand if this is still given by merger and annihilation of bc or some other mechanism. 

While both in this work and in \cite{Ciccone:2024guw} we have mainly focused on the Dirichlet boundary condition, the next crucial step is to study the Neumann case, which is essential for extrapolating to flat-space physics. The spectrum of glueballs, mesons, and baryons in the confining phase is expected to be governed by the extrapolation to flat space of scaling dimensions of boundary gauge-invariant operators. In particular, 
pions correspond to tilt operators for the broken $SU(n_f)_A$ global symmetry at the boundary. 
While extrapolating from a few perturbative orders might look both daunting and naive, it might still lead to reasonable estimates.\footnote{ 
Alternative, non-perturbative approaches making use of the conformal symmetry of the boundary theory are likewise under active investigation \cite{LoParco:2025Bootstrap}.} It might also be possible that certain parametric limits, such as the 't Hooft planar limit, render the determination of the spectrum more tractable already at finite $L$.
Related to that, it would be useful to explore the large-order behaviour of perturbative series in AdS, in particular of boundary conformal data.\footnote{
Note that, if such series turn out to be Borel resummable, the determination of few perturbative coefficients might be enough to have a decent estimate of the exact result.} For example, it would be interesting to establish the nature of the series \eqref{eq:AllOrderChiSB} and see if, at least in principle, we could reconstruct 
the non-perturbative chiral symmetry breaking in flat space from perturbation theory.

Another promising avenue of investigation would be the study of extended operators, with or without matter, and their fate in the flat-space limit. 
In pure YM with Dirichlet bc for the gauge fields, it is possible to have Wilson lines with endpoints on the boundary, as well as 't Hooft lines with support only on the boundary (for Neumann bc, the behavior of Wilson and 't Hooft lines is exchanged). Although confinement in AdS cannot be detected by the scaling of the expectation value of Wilson loops, we expect it to still be captured in the details of correlation functions of extended operators and of their endpoints.\footnote{A particular class of boundary conformal defects for pure Yang-Mills in AdS$_3$, corresponding to a long confining string in the bulk, has recently been studied in \cite{Gabai:2025hwf}.}

Finally, understanding the nature of the Dirichlet$^*$ bc is still an open issue, which we did not touch upon in this work and certainly deserves further investigation.

\section*{Acknowledgments}
We thank Ankur, Davide Bason, Dean Carmi, Christian Copetti, Victor Gorbenko, Ziming Ji, Shota Komatsu, Stefanos R. Kousvos, Manuel Loparco, Marco Meineri, Alessandro Piazza, Veronica Sacchi, Volodia Schaub and Alessandro Vichi for discussions and/or collaboration on related topics. We also thank I. Jack and H. Osborn for an e-mail correspondence. 
RC also gratefully acknowledges the CERN TH Department and INFN Trieste for hospitality while part of this work was being carried out. Work partially supported by INFN Iniziativa Specifica ST\&FI. The work of RC is supported by the Israel Science Foundation (ISF) grant no. 1487/21, and by the MOST NSF/BSF Physics grant no. 2022726. The work of FDC is supported by the Italian Ministry of University and Research (MUR) under the FIS grant BootBeyond (CUP: D53C24005470001).

\bibliographystyle{JHEP}
\bibliography{QCDInAdS4}

\end{document}